\documentclass[fleqn,usenatbib,useAMS]{mnras}

\DeclareRobustCommand{\VAN}[3]{#2}
\let\VANthebibliography\thebibliography
\def\thebibliography{\DeclareRobustCommand{\VAN}[3]{##3}\VANthebibliography}

\usepackage{amsmath,amsfonts,amssymb} 
\usepackage{bm} 
\usepackage[table,  dvipsnames]{xcolor} 
\usepackage{hyperref} 
\usepackage{nicefrac} 
\usepackage{placeins}

\usepackage[varg]{txfonts} 
\usepackage[T1]{fontenc}

\usepackage{ulem}

\hypersetup{
	pdftitle = {AMF and radiation transport},
	pdfauthor = {Alexandros Ziampras, Richard P. Nelson, Roman R. Rafikov},
	pdfsubject = {},
	colorlinks = {true},
	linkcolor = [rgb]{0,0.35,0.7},
	citecolor = [rgb]{0,0.35,0.7},
	filecolor = [rgb]{0.61,0,0},
	urlcolor = [rgb]{0.61,0,0},
}

\usepackage{graphicx}  
\graphicspath{{figures/}}
\newcommand*{\vcenteredhbox}[1]{\begingroup
\setbox0=\hbox{#1}\parbox{\wd0}{\box0}\endgroup}

\newcommand{\tensorGR}[1]{\overline{\bm{{#1}}}}
\newcommand{\DP}[2]{\frac{\partial{#1}}{\partial{#2}}}
\newcommand{\D}[2]{\frac{\text{d}{#1}}{\text{d}{#2}}}

\newcommand{\G}{\text{G}}
\newcommand{\Mstar}{M_\star}
\newcommand{\Lstar}{L_\star}
\newcommand{\Rp}{R_\mathrm{p}}
\newcommand{\Mp}{M_\mathrm{p}}
\newcommand{\hp}{h_\mathrm{p}}
\newcommand{\Tp}{T_\mathrm{p}}
\newcommand{\Tb}{T_\mathrm{b}}
\newcommand{\Mth}{M_\mathrm{th}}
\newcommand{\Msun}{\mathrm{M}_\odot}
\newcommand{\Lsun}{\mathrm{L}_\odot}
\newcommand{\Mjup}{\mathrm{M}_\mathrm{J}}
\newcommand{\Rgas}{\mathcal{R}}
\newcommand{\cs}{c_\mathrm{s}}

\newcommand{\csadb}{c_\mathrm{s}^\mathrm{ad}}
\newcommand{\OmegaK}{\Omega_\mathrm{K}}
\newcommand{\mean}[1]{\langle {#1} \rangle}
\newcommand{\tauR}{\tau_\mathrm{R}}
\newcommand{\tauP}{\tau_\mathrm{P}}
\newcommand{\tauReff}{\tau_\mathrm{R}^\mathrm{eff}}

\newcommand{\taueff}{\tau_\mathrm{eff}}
\newcommand{\kappaR}{\kappa_\mathrm{R}}
\newcommand{\kappaP}{\kappa_\mathrm{P}}
\newcommand{\cv}{c_\mathrm{v}}
\newcommand{\sigmaSB}{\sigma_\mathrm{SB}}
\newcommand{\vel}{\bm{u}}
\newcommand{\FJ}{F_J}

\newcommand{\FJj}{{\FJ}_0}

\newcommand{\tcool}{t_\mathrm{cool}}
\newcommand{\bsurf}{\beta_\mathrm{surf}}
\newcommand{\bmid}{\beta_\mathrm{mid}}
\newcommand{\btot}{\beta_\mathrm{tot}}

\newcommand{\Qvisc}{Q_\mathrm{visc}}
\newcommand{\Qcool}{Q_\mathrm{cool}}
\newcommand{\Qirr}{Q_\mathrm{irr}}
\newcommand{\Qrad}{Q_\mathrm{rad}}
\newcommand{\Qrelax}{Q_\mathrm{relax}}

\defcitealias{miranda-rafikov-2020b}{MR20b}
\newcommand{\mrb}{\citetalias{miranda-rafikov-2020b}}

\title[Planet-induced gaps and rings: in-plane cooling]{Modeling planet-induced gaps and rings in ALMA disks:\\~the role of in-plane radiative diffusion}

\author[A.~Ziampras et al.]{
Alexandros~Ziampras,$^{1}$\thanks{E-mail: a.ziampras@qmul.ac.uk}
Richard~P.~Nelson,$^{1}$
Roman~R.~Rafikov$^{2,3}$
\\
$^{1}$Astronomy Unit, Dept. of Physics and Astronomy, Queen Mary University of London, London E1 4NS, UK\\
$^{2}$DAMTP, University of Cambridge, Wilberforce Road, Cambridge CB3 0WA, UK\\
$^{3}$Institute for Advanced Study, Einstein Drive, Princeton, NJ 08540, USA
}

\date{Accepted XXX. Received YYY; in original form ZZZ}
\pubyear{2023}

\begin{document}
\label{firstpage}
\pagerange{\pageref{firstpage}--\pageref{lastpage}}
\maketitle

\begin{abstract}
	ALMA observations of protoplanetary disks in dust continuum emission reveal a variety of annular structures. Attributing the existence of such features to embedded planets is a popular scenario, supported by studies using hydrodynamical models. Recent work has shown that radiative cooling greatly influences the capability of planet-driven spiral density waves to transport angular momentum, ultimately deciding the number, position, and depth of rings and gaps that a planet can carve in a disk. However, radiation transport has only been treated via local thermal relaxation, not taking into account radiative diffusion along the disk plane. We compare the previous state-of-the-art models of planet--disk interaction with local cooling prescriptions to our new models that include cooling in the vertical direction and radiative diffusion in the plane of the disk, and show that the response of the disk to the induced spiral waves can differ significantly when comparing these two treatments of the disk thermodynamics. We follow up with synthetic emission maps of ALMA systems, and show that our new models reproduce the observations found in the literature better than models with local cooling. We conclude that appropriate treatment of radiation transport is key to constraining the parameter space when interpreting ALMA observations using the planet--disk interaction scenario.
\end{abstract}



\section{Introduction}
\label{sec:introduction}

Annular features are abundant in observations of protoplanetary disks, one of the more famous examples being the HL~Tau system \citep{alma-etal-2015b}. The DSHARP survey \citep{andrews-etal-2018} has further revealed a multitude of rings and gaps in several systems using dust continuum emission observations \citep{huang-etal-2018}. This provides theorists with high spatial- and spectral-resolution data, useful in constructing realistic physical models that can constrain properties of these systems such as dust opacities \citep[e.g.,][]{birnstiel-etal-2018}, the level of turbulence \citep[e.g.,][]{dullemond-etal-2018}, as well as the masses of planets that they might harbor \citep[e.g.,][]{zhang-etal-2018, zhang-etal-2021}.

The planet--disk interaction scenario is particularly exciting given the detection of a pair of young planets embedded in the disk around PDS~70 \citep{keppler-etal-2018,haffert-etal-2019}, cementing the concept that planets grow in protoplanetary disks. Planets interact with the gaseous disk they are embedded in by launching spiral density waves \citep{ogilvie-lubow-2002,rafikov-2002a} which then steepen into shocks, depositing angular momentum into the gas non-locally and driving the opening of one or more gaps in the process \citep{rafikov-2002b}. The resulting radial gas structure consists of one or more pressure maxima adjacent to such gaps, which can trap dust grains and form the bright rings that are observed in continuum emission. As a result, planet--disk interaction has the potential to explain the ring and gap structures in such disks \citep[e.g.,][]{zhang-etal-2018,guzman-etal-2018}, as well as the non-axisymmetric features such as vortices \citep[e.g.,][]{Koller2003,rafikov-cimerman-2023} or material trapped around the planet's Lagrange points \citep[e.g.,][]{rodenkirch-etal-2021, garrido-deutelmoser-etal-2023}.

Planet-driven gap opening is a process highly sensitive to the thermodynamics of the disk environment. Spiral shocks can heat up the surrounding disk \citep{rafikov-2016,ziampras-etal-2020a}, possibly changing the shape of such spirals \citep[e.g.,][]{zhu-etal-2015} when compared to a locally isothermal approach, where the thermal feedback of spirals on the disk is omitted. At the same time, \citet{miranda-rafikov-2020a} showed that introducing a finite cooling timescale $\tcool=\beta\Omega^{-1}$ as a fraction of the orbital timescale $\OmegaK^{-1}$ subjects planetary spirals to linear radiative damping  concurrently with their nonlinear dissipation through shocking \citep{goodman-rafikov-2001}. This effectively reduces the amount of wave angular momentum that is transported further in the disk, thus weakening the capability of a single planet to open multiple gaps. In complementary studies to the above, \citet{zhang-zhu-2020} confirmed that radiative damping of spiral arms is strongest for $\beta\sim 1$, while \citet{ziampras-etal-2020b} showed that hydrodynamical simulations with radiative cooling could produce fewer rings and gaps in the gas component of realistic ALMA disks compared to locally isothermal models \citep[e.g.,][]{zhang-etal-2018}.

The above studies considered a cooling timescale motivated by the thermal emission of dust grains off of the top and bottom surfaces of the disk. While this is typically the dominant mechanism that sets the temperature profile of the disk \citep{chiang-goldreich-1997}, the diffusion of thermal energy across the disk midplane can be equally or more important around planet-driven spirals due to the sharp temperature gradients around their crests \citep{goodman-rafikov-2001}. \citet{miranda-rafikov-2020b} (hereafter \citetalias{miranda-rafikov-2020b}) explored the role of such in-plane cooling and showed that it may dramatically affect the angular momentum budget of spiral waves, potentially recovering a picture closer to the locally isothermal limit regarding planet-driven rings and gaps.

Their study, while supporting the idea that a single planet could, after all, be responsible for multiple rings and gaps, relied on a simple local model to approximate the effects of in-plane radiation transport. Instead of solving for the diffusion of heat along the disk and across spiral shocks directly, \mrb~treated in-plane cooling as an additional effective local cooling channel, similar to the surface cooling described above. Therefore, the details of radiative diffusion across spiral arms, which could affect their angular momentum budget, were not explicitly captured in that work.

In this study, we combine the cooling models of \citet{ziampras-etal-2020b} with the methodology of \citet{miranda-rafikov-2020a,miranda-rafikov-2020b}, while also including a direct treatment of in-plane radiative diffusion according to \citet{ziampras-etal-2020a} (see Appendix~D therein). This allows us to test different prescriptions for handling disk thermodynamics. In doing so, we aim to compare the above findings with those by \citet{zhang-zhu-2020} and the locally isothermal models of ALMA disks by \citet{zhang-etal-2018}, and investigate whether a more realistic treatment of thermodynamics could help recover the ``single planet, multiple gaps'' scenario.

Our physical framework is described in Sect.~\ref{sec:physics-motivation}. Our methods and results are model-dependent, and are presented in Sects.~\ref{sec:amf-study}~\&~\ref{sec:real-systems}. We discuss our findings in Sect.~\ref{sec:discussion}, and conclude in Sect.~\ref{sec:conclusions}.

\section{Physics and motivation}
\label{sec:physics-motivation}

In this section we lay out our physical framework. We describe the radiative mechanisms that we consider in our models, and motivate our approach.

\subsection{Protoplanetary disk hydrodynamics}
\label{sub:ppd-physics}

We consider a disk of ideal gas with adiabatic index $\gamma$ and mean molecular weight $\mu$ orbiting around a star with mass $M_\star$ and luminosity $\Lstar$. In a vertically integrated framework the gas has a surface density $\Sigma$, specific internal energy $e$, and velocity field $\vel=(u_R, u_\phi)$. The Navier--Stokes equations then read
\begin{subequations}
	\label{eq:navier-stokes}
	\begin{align}
		\label{eq:navier-stokes-1}
		\DP{\Sigma}{t} + \vel\cdot\nabla\Sigma=-\Sigma\nabla\cdot\vel,
	\end{align}
	\begin{align}
	\label{eq:navier-stokes-2}
		\Sigma\DP{\vel}{t}+ \Sigma(\vel\cdot\nabla)\vel=-\nabla P -\Sigma\nabla(\Phi_\star+\Phi_\mathrm{p}) +\nabla\cdot\bm{\tensorGR{\sigma}},
	\end{align}
	\begin{align}
	\label{eq:navier-stokes-3}
		\DP{\Sigma e}{t} + \vel\cdot\nabla(\Sigma e)=-\gamma\Sigma e\nabla\cdot\vel+Q
	\end{align}
\end{subequations}
The stellar potential at distance $R$ is given by $\Phi_\star = -\G\Mstar/R$, the viscous stress tensor is denoted by $\tensorGR{\sigma}$, and the vertically integrated pressure $P=(\gamma-1)\Sigma e$ follows the ideal gas equation of state. The isothermal sound speed is then $\cs=\sqrt{P/\Sigma}=\sqrt{\Rgas T/\mu}$, with $T$ being the (vertically constant) gas temperature, and relates to the adiabatic sound speed as $\csadb = \cs\sqrt{\gamma}$.
We can also define the pressure scale height $H=\cs/\OmegaK$, where $\OmegaK = \sqrt{\G M_\star/R^3}$ is the Keplerian angular velocity, and the aspect ratio $h = H/R$. In the above, $\G$ and $\Rgas$ denote the gravitational constant and ideal gas constant, respectively. The term $Q$ encompasses any additional sources or sinks of thermal energy.

Our models include an embedded planet with mass $\Mp$, fixed at $R=\Rp$ in the form of a potential term in Eq.~\eqref{eq:navier-stokes-2}. The planetary potential follows a Plummer prescription 
\begin{equation}
	\Phi_\text{p} = -\frac{\G \Mp}{\sqrt{d^2 + \epsilon^2}},\qquad \bm{d} = \bm{R} - \bm{R}_\text{p}
\end{equation}
with a smoothing length $\epsilon=0.6\,H_\text{p}$ that aims to account for the vertical stratification of the disk, similar to \citet{mueller-etal-2012}. The planet is not allowed to migrate or accrete.

Finally, our models consider a reference frame centered on the star. While the indirect term by the star--planet system orbiting about their center of mass is included, we do not account for the disk's feedback on either object.

\subsection{Radiative effects}
\label{sub:radiative-effects}

The thermal structure of the disk is determined by a number of different physical processes represented by the terms in Eq.~\eqref{eq:navier-stokes-3}. In this study, we consider the following radiative source and sink terms.

\subsubsection{Viscous heating}

Following the $\alpha$-viscosity model of \citet{shakura-sunyaev-1973}, the gas kinematic viscosity is $\nu=\alpha\csadb H$. We can then write:
	\begin{equation}
		Q_\text{visc} = \frac{1}{2\nu\Sigma}\mathrm{Tr}(\tensorGR{\sigma}^2) \approx \frac{9}{4}\nu\Sigma\OmegaK^2.
	\end{equation}
	In this study, we explore the nearly-inviscid regime with $\alpha=10^{-5}$. This allows our disk to essentially not accrete and remain passively heated by stellar irradiation, while minimizing the effects of numerical diffusion. Furthermore, for this low $\alpha$, viscous damping of the angular momentum flux carried by spiral arms is negligible \citepalias{miranda-rafikov-2020b}.

\subsubsection{Vertical cooling}

The gas can transfer its thermal energy to nearby dust grains which then emit as a black body, cooling the disk in the process. Following \citet{menou-goodman-2004}, we have
	\begin{equation}
		\label{eq:Qcool}
		Q_\text{cool} = -2\sigmaSB\frac{T^4}{\taueff},\qquad\taueff = \frac{3\tauR}{8} + \frac{\sqrt{3}}{4} + \frac{1}{4\tauP},
	\end{equation}
	%
	where $\taueff$ is the effective optical depth according to \citet{hubeny-1990}, and $\tau_{\mathrm{R,P}}=\frac{1}{2}\kappa_{\mathrm{R,P}}\Sigma$ is the optical depth at the disk midplane for Rosseland and Planck mean opacities $\kappaR$ and $\kappaP$, respectively. In our study, and in the analysis below, we assume that $\kappa_{\mathrm{P}}=\kappa_{\mathrm{R}}$ and $\tauR=\tauP=\tau$.

\subsubsection{Stellar irradiation}

The disk intercepts part of the incoming radiation by the star. Similar to \citet{menou-goodman-2004} we can approximate this effect as
	\begin{equation}
		\label{eq:Qirr}
		Q_\text{irr} = 2\frac{\Lstar}{4\pi R^2} (1-\epsilon)\frac{\theta}{\taueff},
	\end{equation}
	where we adopt $\epsilon=0.5$ for the disk albedo, $\theta$ is the flaring angle, and $\taueff$ is the effective optical depth similar to Eq.~\eqref{eq:Qcool}. In principle, $\theta$ is radius-dependent \citep[see e.g.,][]{chiang-goldreich-1997,menou-goodman-2004}, but constant small values of the order of $\theta=0.02$--0.03 are often used in the literature \citep[e.g.,][]{zhang-etal-2018}. By equating $Q_\text{irr}=Q_\text{cool}$ and assuming $\kappaR=\kappaP$, one finds the well-known passively-heated disk temperature profiles $T(R)\propto R^{-0.5}$ for $\theta=\text{const.}$, or $T(R)\propto R^{-3/7}$ for $\theta = R\frac{\text{d}h}{\text{d}R}$.

\subsubsection{Radiative diffusion}

Dust grains emit in all directions, and as a result a local perturbation of temperature drives a radiative flux not only vertically but also through the disk midplane. We model this process using the flux-limited diffusion approximation \citep{levermore-pomraning-1981}, by writing
	\begin{equation}
		\label{eq:Qrad}
		Q_\text{rad} = -\sqrt{2\pi}H\,\nabla\cdot\bm\left(\lambda\frac{4\sigmaSB}{\kappaR\rho_\text{mid}}\nabla T^4\right).
	\end{equation}
	Here, $\cv=\frac{\Rgas}{\mu(\gamma-1)}$ is the heat capacity of the gas at constant volume, $\rho_\text{mid} = \frac{1}{\sqrt{2\pi}} \frac{\Sigma}{H}$ is the volume density at the disk midplane, and $\lambda$ is a flux limiter following \citet{kley-1989}:
	\begin{equation}
		\lambda(R^\prime) = \left\{\begin{array}{lr}
		\frac{2}{3+\sqrt{9+10 {R^\prime}^2}}, & 0\leq R^\prime\leq 2\\
		\frac{10}{10R^\prime + 9 + \sqrt{180R^\prime+ 81}}, & 2\leq R^\prime\leq \infty.
		\end{array}\right.,\quad R^\prime \equiv 4l_\text{rad}\frac{|\nabla T|}{T},
	\end{equation}
	where $l_\text{rad} = 1/(\kappaR\rho_\text{mid})$ is the photon mean free path. This flux limiter handles the transition from the optically thick, diffusion limit ($\lambda\rightarrow1/3$) to the optically thin, free-streaming limit ($\lambda\rightarrow1/R^\prime$).


\subsection{Implementing radiative effects}
\label{sub:implementing-radiative-effects}

With the above in mind, there are (at least) 4 different recipes of handling gas thermodynamics in our framework. From simplest to most complex:
\begin{enumerate}
	\item \textbf{locally isothermal}: the energy equation is dominated by radiative processes ($Q$ in Eq.~\eqref{eq:navier-stokes-3}) such that any excess cooling/heating due to (de)compression is removed instantaneously and the gas is always in thermal equilibrium. The energy equation is then omitted. Instead, we assume that $T(R)$ is time-independent and given by a balance between $Q_\text{irr}$ (heating) and $Q_\text{cool}$ (cooling), computed \textit{a priori}.
	\item \textbf{adiabatic}: changes in temperature solely depend on gas (de)compression (first term on RHS of Eq.~\eqref{eq:navier-stokes-3}), and no radiative effects are considered; entropy is conserved in the absence of shocks.
	\item \textbf{surface cooling}: the gas can heat up via shock heating, $Q_\text{irr}$, and $\Qvisc$, and cool through the disk surfaces via $Q_\text{cool}$. In our vertically-integrated models, this cooling mechanism is local.
	\item \textbf{fully radiative}: in-plane radiative diffusion is added to the above by including $Q_\text{rad}$.
\end{enumerate}

It is quite common to account for the mechanisms outlined above using a parametrized approach. Assuming the gas cools over a cooling timescale $\tcool$, we define the dimensionless cooling timescale $\beta=\tcool\OmegaK$. This $\beta$ can be a constant, allowing for scale-free models, or a function of one or more physical quantities, allowing for more realistic models that might explore certain mechanisms or target specific observations. In that case, one can assume that there exists an equilibrium temperature profile $T_0(R)$ that remains largely unchanged in time (e.g., as the result of a balance between $Q_\text{irr}$ and $Q_\text{cool}$), and then remove excess heat from the disk over a timescale $\beta$ such that
\begin{equation}
	\label{eq:Qrelax}
	Q_\text{relax} = -\Sigma\cv\frac{T-T_0}{\beta}\OmegaK.
\end{equation}
This approach allows the disk temperature profile to remain reasonably close to $T_0$ while also accounting for the interplay between compression heating and radiative terms. This is useful in studying the radiative damping of processes such as the spiral arms of a planet \citep[e.g.,][]{zhang-zhu-2020,miranda-rafikov-2020a,miranda-rafikov-2020b}, the opening of a circumbinary cavity \citep[e.g.,][]{sudarshan-etal-2022,pierens-nelson-2022}, the development of various instabilities \citep[e.g.,][]{bethune-latter-2021,pfeil-klahr-2021}, etc.

Equations~\eqref{eq:Qcool}~\&~\eqref{eq:Qrad} allow us to define cooling timescales associated with surface (vertical) and in-plane cooling respectively. Following \citet{ziampras-etal-2020b}, the surface cooling timescale $\bsurf$ can be written as 
\begin{equation}
	\label{eq:bsurf}
	\bsurf = \frac{\Sigma e}{|Q_\text{cool}|} \OmegaK.
\end{equation}

Treatment of the in-plane cooling is not as simple, because heat diffuses through the midplane on a timescale that depends on the diffusion lengthscale $l_\text{d}$. Taking this into account, \citetalias{miranda-rafikov-2020b} argued that the in-plane cooling of the $m$-th azimuthal Fourier component of the thermal perturbation can be approximated via a $\beta$-cooling term with the cooling timescale $\bmid^m$ given by
\begin{equation}
	\label{eq:bmid}
	\bmid^m \approx \frac{\OmegaK}{\eta}\left[(l_\text{d}^m)^2 + \frac{l_\text{rad}^2}{3}\right],\qquad \eta=\frac{16\sigmaSB T^3}{3\kappaR\rho_\text{mid}^2 \cv},
\end{equation}
where the two terms in the square brackets represent the characteristic photon diffusion lengthscales in the optically thick and thin limits, respectively. Here, $\eta$ is a radiative diffusion coefficient associated with Eq.~\eqref{eq:Qrad}. \citetalias{miranda-rafikov-2020b}  further suggested that the cooling effect of many Fourier perturbation harmonics can be reduced approximately to a local cooling term with a single $\beta_\star$ given by Eq.~\eqref{eq:bmid} with $l_\text{d}^m$ replaced with $l_\text{d}^\star = 1/k_\star$, such that
\begin{equation}
	\label{eq:kstar}
	k_\star \approx \frac{1}{l_\text{rad}}\left(\frac{\pi}{2}\right)^{1/2}\left|\frac{\Omega_\mathrm{p}}{\Omega}-1\right|\frac{m_\star}{\tau},\quad m_\star \approx \frac{1}{2 \hp}.
\end{equation}

In our present study, we simplify this local prescription even further by assuming that the dominant diffusion lengthscale is $l_\text{d}^\star\approx H$ \citep[e.g.,][]{flock-etal-2017}, and write
\begin{equation}
	\label{eq:bmidH}
	\bmid \approx \frac{\OmegaK}{\eta}\left(H^2 +  \frac{l_\text{rad}^2}{3}\right).
\end{equation}
We note that the dominant diffusion lengthscale will be smaller than $H$ near shocks. As a result, this approach is only an approximation and cannot allow a direct comparison with the results of \mrb. We nevertheless carry out a comparison in Appendix~\ref{apdx:mr20b}, finding that the recipe of \citetalias{miranda-rafikov-2020b} does not always capture the the behavior of the spiral AMF appropriately.

Finally, we can combine the two timescales $\bsurf$ and $\bmid$ into an effective cooling timescale $\btot$ that accounts for both vertical and in-plane cooling. Following \citetalias{miranda-rafikov-2020b}, we write
\begin{equation}
	\label{eq:btot}
	\btot^{-1} = \bsurf^{-1} + \bmid^{-1}.
\end{equation}  
With some algebra, we can show using Eqs.~\eqref{eq:Qcool}, \eqref{eq:bsurf}, \eqref{eq:bmid}, \eqref{eq:bmidH} that in our approach with $l_\text{d}=H$, $\btot$ can then be written as
\begin{equation}
	\label{eq:factor-f}
	\btot = \frac{\bsurf}{1+f},\qquad f = \frac{\bsurf}{\bmid} = 16\pi\frac{\tau\,\tau_\text{eff}}{6\tau^2+\pi},
\end{equation}
implying that models that do not account for in-plane cooling will considerably overestimate the cooling timescale, as also shown by \citetalias{miranda-rafikov-2020b}.

\begin{figure}
	\includegraphics[width=\columnwidth]{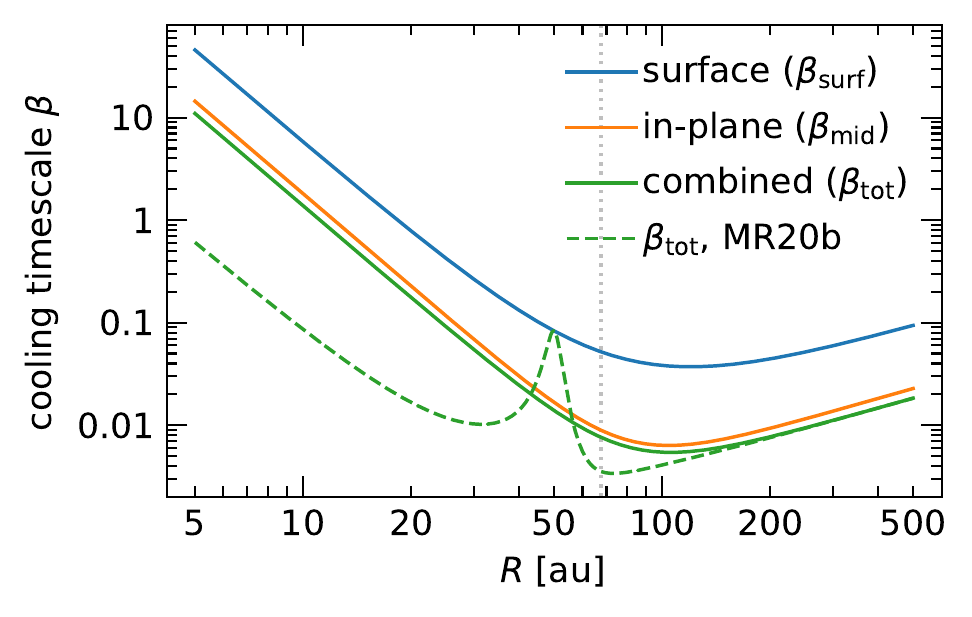}
	\caption{Estimates of the cooling timescale using Eqs.~\eqref{eq:bsurf},~\eqref{eq:bmid}~\&~\eqref{eq:btot} for an example system similar to the setup of \citetalias{miranda-rafikov-2020b}. The in-plane cooling timescale $\bmid$ is about 3--4 times lower than the surface cooling time $\bsurf$, highlighting the importance of in-plane cooling. Here, we chose $\Mstar=\Msun$, $R_0=50$\,au, $\mu=2$, $h_0=0.1$, $\Sigma_0=10$\,g/cm${}^2$, $\Sigma\propto R^{-1}$, $T\propto R^{-1/2}$, and $\kappa = 2\times10^{-4} T^2$. The solid green line assumes $l_\text{d} = H$ (Eq.~\eqref{eq:bmidH}), while dashed green uses the recipe by \citetalias{miranda-rafikov-2020b} with $l_\text{d}\approx 1/k_\star$ (Eq.~\eqref{eq:kstar}) and $\Rp=50$\,au.}
	\label{fig:example-beta}
\end{figure}

An example of $\beta$ profiles for a passively-heated disk mimicking the setup of \citetalias{miranda-rafikov-2020b} is shown in Fig.~\ref{fig:example-beta}. As they showed, $\bmid$ is approximately 3--4 times lower than $\bsurf$, pushing $\btot$ to $\approx$0.2--0.25\,$\bsurf$. This highlights the necessity of treating in-plane cooling, which \citetalias{miranda-rafikov-2020b} demonstrated in their study.

However, a caveat of implementing in-plane cooling with this $\beta$ approach is that it is inherently localized: a local increase in temperature should result in both surface cooling, which cools the hot spot over time depending on $\bsurf$, as well as radiative diffusion. The latter spreads the heat to neighboring areas, cooling the hot spot over $\bmid$ but also heating up its surroundings. This last heating effect is missed when using $\btot$ as a proxy for in-plane cooling, as this approach always relaxes the gas towards the equilibrium state $T_0$. This can result in significant differences around regions of sharp temperature gradients (such as shocks) between a $\btot$-cooled model and one that explicitly accounts for $\Qrad$.

In the context of planet--disk interaction, \citetalias{miranda-rafikov-2020b} have already shown that in-plane cooling through $\btot$ can affect gap opening by a planet embedded in a disk by essentially allowing the gas to cool through the midplane. This approach, however, cannot capture the heating due to thermal diffusion around a spiral shock, which could further affect the gap opening process. For that, directly accounting for $Q_\text{rad}$ is required.

\begin{figure}
	\includegraphics[width=\columnwidth]{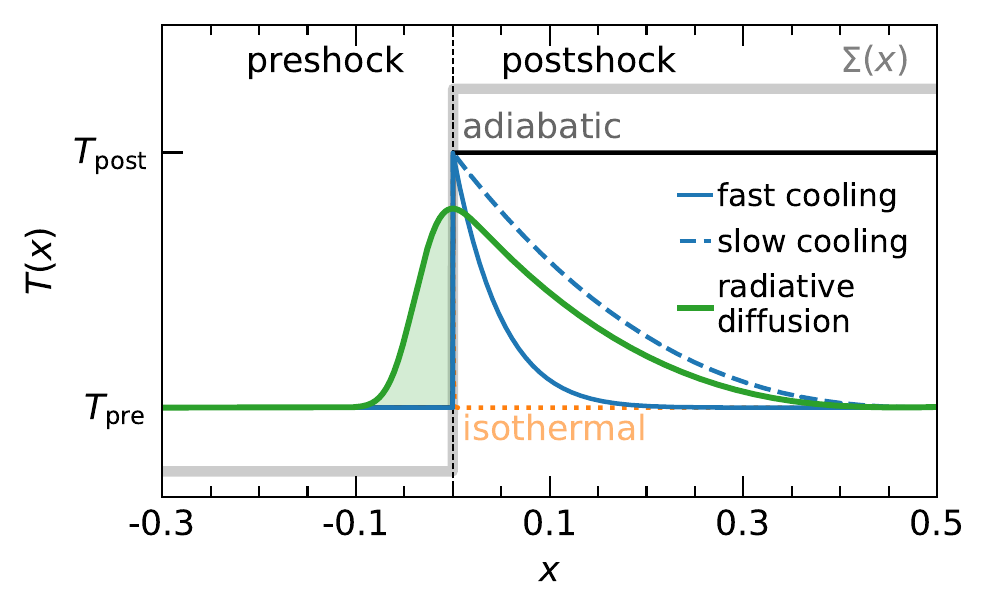}
	\caption{Cartoon example of the temperature profile around a one-dimensional shock front (at $x=0$). The higher post-shock temperature $T_\text{post}$ can relax to the pre-shock temperature $T_\text{pre}$ at a rate that depends on the cooling timescale $\beta$ (faster for smaller $\beta$). Accounting for radiative diffusion further smears the shock front, resulting in a heating of the pre-shock region as well. Of course, spiral shocks are more complex than this example, with $\Sigma$ and $T$ reverting to their pre-shock values after a few $H$ (see Fig.~\ref{fig:rad-corr-shock}).}
	\label{fig:shock-front}
\end{figure}

To illustrate our point, we consider a one-dimensional shock front in Fig.~\ref{fig:shock-front}. Accounting for surface cooling relaxes the hot, post-shock region to its initial, pre-shock state over a lengthscale that depends on the cooling timescale (blue curves). By treating in-plane cooling as a local effect (i.e., using $\bmid$) we essentially enable an additional cooling pathway for the gas, reducing the effective $\beta$ from $\bsurf$ to $\btot$. However, when handling in-plane cooling via radiative diffusion, the temperature spike at the shock front spreads into both the pre- and post-shock regions (green curve). Considering how the deposition of angular momentum by planet-driven spiral arms and the subsequent gap opening are very sensitive functions of the radiative damping of such shocks, it becomes clear that a more realistic treatment of the thermodynamics could have an impact on planet--disk interaction in this context.

With fully radiative models (i.e., accounting for $\Qrad$), we aim to explore the interplay between surface and in-plane cooling and combine the findings of \citet{miranda-rafikov-2020a,miranda-rafikov-2020b}, \citet{zhang-zhu-2020}, and \citet{ziampras-etal-2020b} into one complete, coherent picture.

We investigate two different scenarios: we start with a wide parameter study that aims to quantify the planet's gap opening capacity with a suite of short-term, high resolution numerical models, presented in Sect.~\ref{sec:amf-study}. We then use our models to identify different regimes of cooling, and demonstrate the effect of in-plane radiative diffusion with long-term simulations and simplistic synthetic observations of continuum emission for two ALMA systems, AS~209 and Elias~20, in Sect.~\ref{sec:real-systems}. The specifics of the numerical setup for each set of models are covered in their respective sections.

\section{The role of cooling on spiral AMF}
\label{sec:amf-study}

In this section we present a high-resolution, wide parameter study of radiative planet--disk interaction. Our aim is to measure the planet's gap opening capabilities using the angular momentum flux (AMF) carried by the planet-driven spiral arms as a metric of the momentum transport and deposition in the disk. Following \mrb, in the linear approximation we define the spiral AMF as 
\begin{equation}
	\label{eq:amf}
	\FJ(R) = R^2\bar{\Sigma} \oint\limits_\phi \delta u_R\,\delta u_\phi \text{d}\phi,\qquad \delta u_x = u_x - \bar{u}_x,\quad x\in\{R,\phi\}
\end{equation}
where a bar denotes an azimuthally averaged quantity. The AMF can then be normalized to a reference value \citep{goldreich-tremaine-1980}
\begin{equation}
	\label{eq:amf0}
	\FJj = \left(\frac{\Mp}{\Mstar}\right)^2 \hp^{-3} \Sigma_\text{p} R^4 \Omega_\text{p}^2.
\end{equation}

In order to quantify the effects of cooling, both surface and in-plane, we explore the $\bsurf$ parameter space for two different values of the optical depth $\tau$. Varying the cooling timescale $\bsurf$ allows us to study the spiral AMF as the disk transitions from adiabatic ($\bsurf\rightarrow\infty$) to isothermal ($\bsurf\rightarrow0$), while varying the optical depth $\tau$ controls the efficiency of both surface cooling and in-plane radiative diffusion. We then carry out models for the equations of state \texttt{iso}, \texttt{adb}, \texttt{surf}, and \texttt{rad}, listed in Table~\ref{table:eos}.

In the following sections we describe our initial and boundary conditions for each model, and describe our findings.

\subsection{Model setup and methods}
\label{sub:amf-model}

We utilize the numerical hydrodynamics code \texttt{PLUTO} \texttt{v4.4} \citep{mignone-etal-2007} in a 2D cylindrical polar $\{R, \phi\}$ geometry. \texttt{PLUTO} uses a finite-volume Godunov scheme to integrate Eqs.~\eqref{eq:navier-stokes} in time using a Riemann solver. Our computational domain extends radially in the range $R\in[0.4\text{--}2.5]\,R_\mathrm{p}$, with $R_\mathrm{p} = 1\,\text{au}$, and covers the full azimuthal range $\phi\in[0, 2\pi]$.

In our models we also use the FARGO method by \citet{masset-2000}, which has been implemented into \texttt{PLUTO} by \citet{mignone-etal-2012} and yields a speedup factor of $\sim10$ alongside improving the accuracy of the numerical scheme. This is achieved by subtracting the background Keplerian flow before solving the Riemann problem across all cell interfaces, substantially relaxing timestep limitations in the rapidly rotating inner disk.

Regarding our numerical setup, we use second-order accurate reconstruction and time marching schemes (\texttt{LINEAR} and \texttt{RK2} respectively) with the \texttt{hllc} Riemann solver \citep{toro-etal-1994} and the \texttt{VAN\_LEER} slope limiter \citet{vanleer-1974}. To minimize dissipation we make use of a frame corotating with the planet, and activate the option \texttt{CHAR\_LIMITING}. Using a higher-order reconstruction
did not affect our results.

\begin{table*}
	\begin{center}	
	\caption{Models referenced throughout this study. Model \texttt{surf} aims to reproduce \citet{ziampras-etal-2020b}, while \texttt{bsurf} and \texttt{btot} aim to reproduce \citetalias{miranda-rafikov-2020b}. Model pairs \texttt{surf}--\texttt{bsurf} and \texttt{rad}--\texttt{btot} consider the same physics with different implementations (source terms and $\beta$ cooling, respectively). Models \texttt{iso}, \texttt{adb}, \texttt{surf}, and \texttt{rad} are introduced in Sect.~\ref{sub:amf-model} and discussed throughout Sect.~\ref{sec:amf-study}, while models \texttt{bsurf} and \texttt{btot} are discussed in Sect.~\ref{sub:amf-beta-vs-Q}.}
	\label{table:eos}
	\begin{tabular}{c | c | c}
		\hline
		tag & physics & relevant terms and equations \\
		\hline
		\texttt{iso} & locally isothermal (instantaneous cooling) & no energy equation \\
		\texttt{adb} & adiabatic (no cooling) & adiabatic compression:~\eqref{eq:navier-stokes-3} \\
		\hline
		\texttt{surf} & surface cooling only & \eqref{eq:navier-stokes-3}; $\Qirr$:~\eqref{eq:Qirr}, $\Qcool$:~\eqref{eq:Qcool} \\
		\texttt{rad} & surface cooling and in-plane radiative diffusion & \eqref{eq:navier-stokes-3}; $\Qirr$:~\eqref{eq:Qirr}, $\Qcool$:~\eqref{eq:Qcool}, $\Qrad$:~\eqref{eq:Qrad} \\
		\hline
		\texttt{bsurf} & $\beta$ relaxation with surface term & \eqref{eq:navier-stokes-3}; $\Qrelax$:~\eqref{eq:Qrelax}, $\bsurf$:~\eqref{eq:bsurf} \\
		\texttt{btot} & $\beta$ relaxation with surface and in-plane terms & \eqref{eq:navier-stokes-3}; $\Qrelax$:~\eqref{eq:Qrelax}, $\btot$:~\eqref{eq:btot} \\
	\end{tabular}
	\end{center}	
\end{table*}

Throughout this parameter scan we keep the (initial) aspect ratio at the planet's position fixed at $\hp = 0.05$ and the planet's mass at $\Mp=0.3\,\Mth=3.75\times10^{-5}\,\Mstar$. Here, $\Mth \approx \hp^3 \Mstar$ is the disk thermal mass at the planet's location \citep{rafikov-2002a}.

To create an environment as simple as possible, we set $\Sigma$, $\bsurf$ and $\tau$ to be constant throughout the domain. Given the nonlinear dependence of $\bsurf$ and $\tau$ on $\Sigma$, $T$, and $\kappa$, this requires unphysical choices for $\Sigma$ and $\kappa$ for some of our models. Nevertheless, we will later show that our approach can be justified by looking at several ALMA systems observed with the DSHARP survey \citep{andrews-etal-2018} in the $\bsurf$--$\tau$ space. As a result, our choices for $\Sigma$ and $\kappa$ (to control $\bsurf$ and $\tau$) as well as $\Lstar$ (to control $\hp$) can be completely arbitrary.

By keeping $\Sigma$ and $\kappa$ constant, we can have radially constant profiles for $\tau$. By further adopting a temperature profile $T(R)\propto R^{-0.5}$ we can also maintain a constant $\bsurf$. This allows for constant parameters throughout the disk while being consistent with a realistic, passively irradiated disk model. A table with all values of $\Sigma$ and $\kappa$ that we explored is provided in Appendix~\ref{apdx:tables}.

Our fiducial model consists of a disk with $\tau=10$, $\bsurf=1$, $\hp=0.05$, and $\Mp=0.3\,\Mth$. After a resolution analysis, presented in detail in Appendix~\ref{apdx:resolution}, we opt for a resolution of 32 cells per scale height in both directions at the planet's location, for a fiducial grid of $1200\times4096$ cells.

We use a locally isothermal and an adiabatic model (labeled ``\texttt{iso}'' and ``\texttt{adb}'') with $\hp=0.05$ and $\Mp=0.3\,\Mth$ as references. Then, we also run models with surface cooling only (``\texttt{surf}''), which keep only $Q_\mathrm{irr}$ and $Q_\mathrm{cool}$ nonzero, and models with both surface and in-plane cooling (``\texttt{rad}''), which additionally include the in-plane radiative diffusion term $Q_\mathrm{rad}$. These four sets of models, which represent the different recipes listed in Sect.~\ref{sub:implementing-radiative-effects}, are described in more detail in Table~\ref{table:eos}. 

We note here that we use $\bsurf$ as a parameter to control the efficiency of cooling in models \texttt{surf} and \texttt{rad}. These models utilize the radiative terms introduced in Sect.~\ref{sub:radiative-effects} and implemented in Eq.~\eqref{eq:navier-stokes-3}, with disk parameters tailored such that the surface cooling timescale $\bsurf$ (defined via Eq.~\eqref{eq:bsurf}) has a desired value. It is very important to keep in mind that $\bsurf$ is used here only for parametrization of these models and does not assume a $\beta$ cooling prescription, which would implement $\bsurf$ via Eq.~\eqref{eq:Qrelax}. For a comparison with models using $\beta$ cooling, see Sect.~\ref{sub:amf-beta-vs-Q}.

In all models the planet grows over one orbit using the formula by \citet{devalborro-etal-2006}, and each model is run for ten planetary orbits so that the global structure of the spiral AMF and the thermal structure of the disk around spirals are established. In addition, wave damping zones are employed in the radial direction between $R\in[0.4,0.5]\cup[2.1,2.5]\,\Rp$ following \citet{devalborro-etal-2006}, with a damping timescale of 0.1 orbits at the corresponding radial boundary. This ensures that spiral arms are not reflected at the boundary edge, which is otherwise closed.

As we will show in the next section, it is difficult to quantify the differences in radial profiles of $\FJ$ between our models due to the size of our parameter study (over 200 models). For this reason, we define a proxy for the planet's ``capacity to open multiple gaps'' for a given thermodynamic prescription as
\begin{equation}
	\label{eq:factor-G}
	G = \frac{J}{J^\text{adb}},\qquad J = \int\limits_{0.5\,\Rp}^{1.5\,\Rp} \FJ \text{d}R,
\end{equation}
where $J^\text{adb}$ represents the adiabatic case.
Effectively, a high $G$ translates to weak radiative wave damping and a multiple-gap configuration ($\beta\rightarrow0$ or $\beta\rightarrow\infty$), whereas a low $G$ translates to strong wave damping and a single gap around the planet and is expected for $\bsurf\sim1$ \citep[see Fig.~4 in][]{miranda-rafikov-2020a}. We choose these radial limits to cover the entire inner disk, but exclude the outskirts of the disk to avoid some oscillations in the radial profiles of the AMF (see e.g., Fig.~\ref{fig:fiducial-amf}) which slightly polluted our results without changing their general behavior.

\subsection{Results}
\label{sub:amf-results}

In this section we present our results for the models described in Sect~\ref{sub:amf-model} in the context of spiral AMF.

\subsubsection{Fiducial model}

\begin{figure}
	\includegraphics[width=\columnwidth]{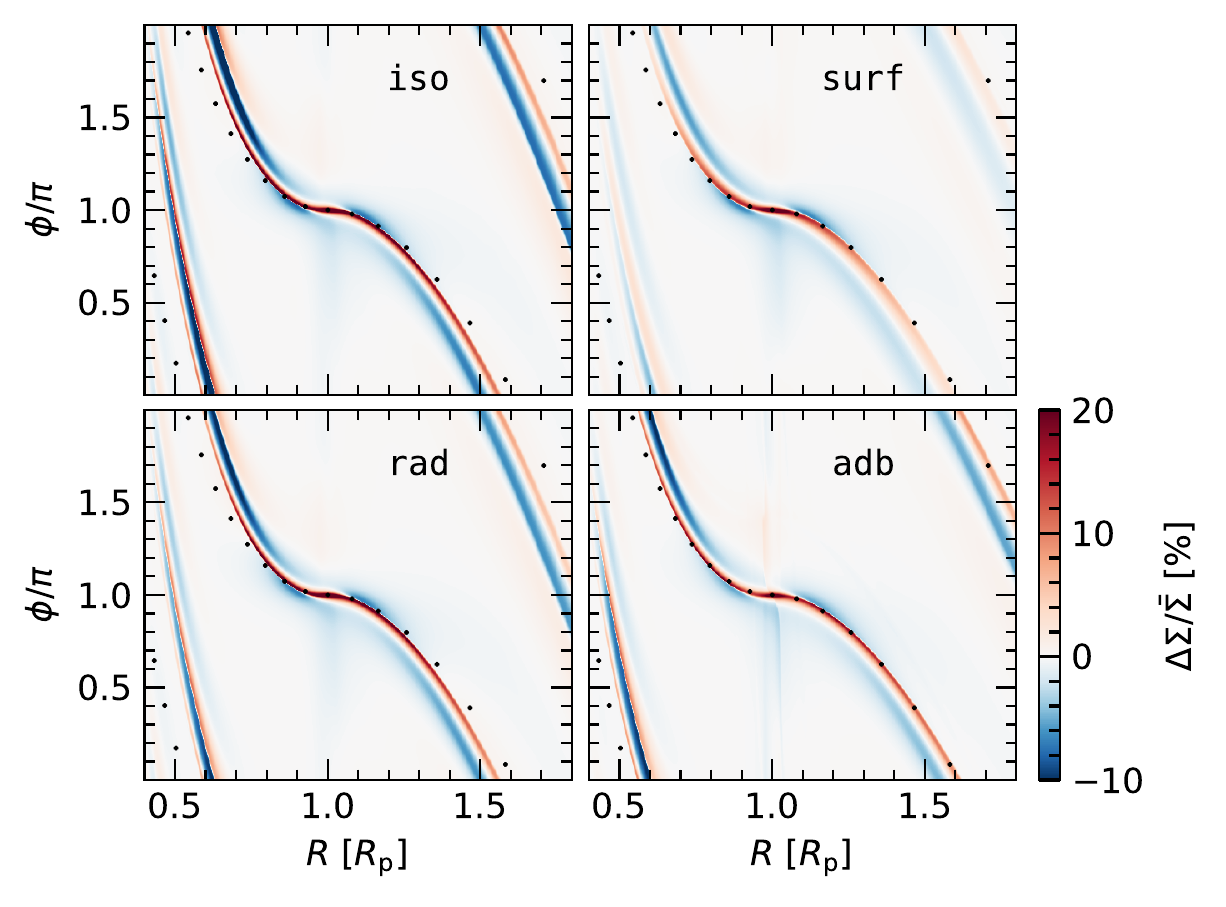}
	\caption{Two-dimensional heatmaps of surface density perturbations in our fiducial model for various thermodynamic treatments. Planet-driven spiral arms permeate the disk, with minor deviations around the planet's corotating region as well. Model \texttt{surf} shows fainter spirals due to the effects of surface cooling, while accounting for in-plane radiation transport (model \texttt{rad}) helps recover a more isothermal-like picture. Dots help guide the eye, showing the different propagation speed of spirals in different panels.} 
	\label{fig:fiducial-planet-sig}
\end{figure}

\begin{figure}
	\includegraphics[width=.99\columnwidth]{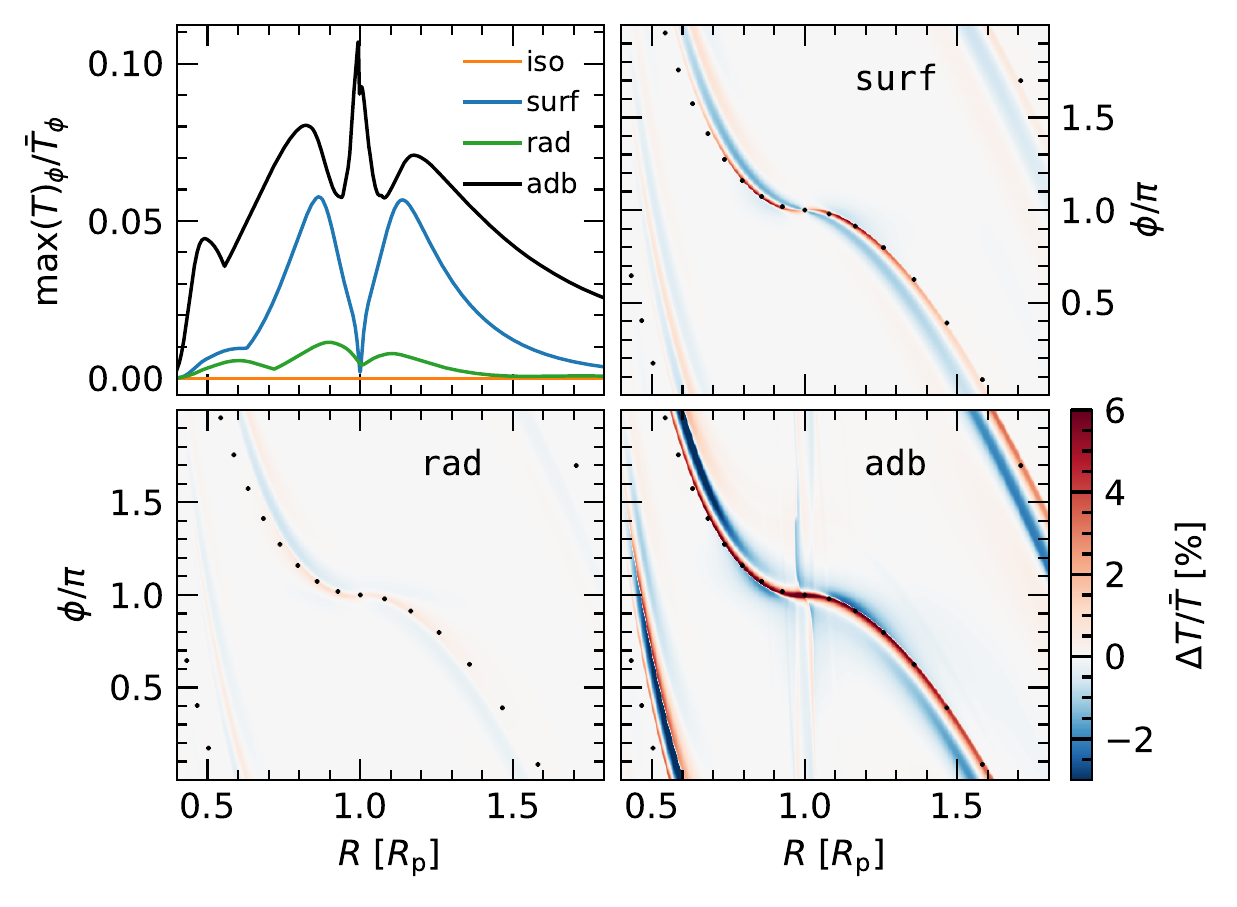}
	\caption{Temperature deviations plotted similar to Fig.~\ref{fig:fiducial-planet-sig}. Depending on the rate of cooling, the temperature contrast around spiral arms can be very weak (\texttt{rad}) to very strong (\texttt{adb}). The top left panel shows the maximal contrast at a given radius to quantify these differences. Temperature spikes near $R=0.5$--0.6 in the inner disk exist due to the excitation of secondary spiral shocks, while a spike at $\Rp$ in the adiabatic model corresponds to hot material in the planet's Hill sphere.}
	\label{fig:fiducial-planet-tmp}
\end{figure}

We start by looking at the global structure of the disk for our models with $\tau=10$, $\bsurf=1$, $\hp=0.05$, $\Mp=0.3\,\Mth$ at $t=10$~orbits to provide some context. Figures~\ref{fig:fiducial-planet-sig}~\&~\ref{fig:fiducial-planet-tmp} show the global deviations in surface density $\Sigma$ and temperature $T$ respectively, after subtracting the azimuthally averaged disk profile.

As expected, these deviations are localized around the planet-driven spiral arms, with minor features inside the planet's corotating region. In particular, model \texttt{surf} shows weaker spiral arms compared to other models (Fig.~\ref{fig:fiducial-planet-sig}), while spirals in model \texttt{rad} are significantly more diffuse in terms of temperature contrast compared to those in models \texttt{surf} or \texttt{adb} (Fig.~\ref{fig:fiducial-planet-tmp}).

The latter is to be expected: the model with in-plane radiation transport allows diffusion of heat along the midplane, which smears temperature gradients about the crests of spiral arms. In addition, it has an effectively shorter cooling timescale since the gas can cool via both the vertical and in-plane channel. As a result, model \texttt{rad} behaves more akin to a locally isothermal model---where no temperature deviations are observed---as far as temperature is concerned.

\begin{figure}
	\includegraphics[width=\columnwidth]{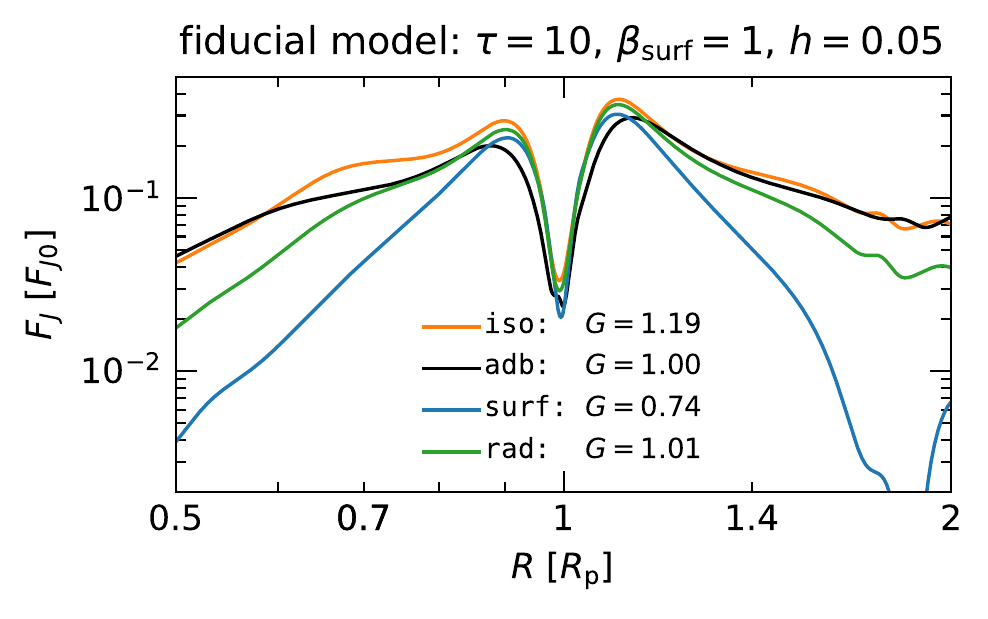}
	\caption{Normalized angular momentum flux (AMF) $\FJ$ due to spiral arms for our fiducial model as a function of equation of state. Values of the metric $G$ defined by Eq.~\eqref{eq:factor-G} are indicated for each model.  We see a distinctly lower AMF for model \texttt{surf}, consistent with previous findings given that $\bsurf=1$.}
	\label{fig:fiducial-amf}
\end{figure}

We then plot radial profiles of the normalized spiral AMF $\FJ/\FJj$ for these four different thermodynamic assumptions in Fig.~\ref{fig:fiducial-amf}, also showing for illustration the values of $G$ defined by Eq.~\eqref{eq:factor-G} for each model. Since $\bsurf=1$, the AMF in model \texttt{surf} is weaker than either the isothermal or adiabatic limits. At the same time, we find a considerably higher $\FJ$ in model \texttt{rad} compared to \texttt{surf} for $\bsurf=1$, consistent with the findings of \citetalias{miranda-rafikov-2020b}.

\FloatBarrier

\subsubsection{Dependence of spiral AMF on $\bsurf$}
\label{sub:amf-beta}

\begin{figure}
	\includegraphics[width=\columnwidth]{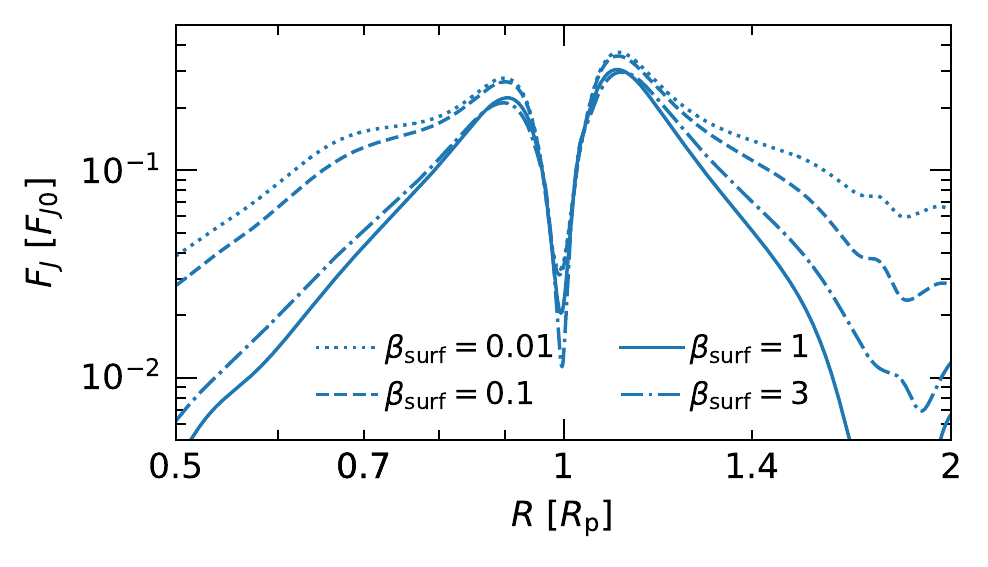}
	\caption{Dependence of the spiral AMF on the surface cooling timescale $\bsurf$ for model \texttt{surf}, similar to Fig.~\ref{fig:fiducial-amf}. We find a minimum in $\FJ$ for $\bsurf\approx1$, consistent with previous reports \citep{miranda-rafikov-2020a}.}
	\label{fig:amf-beta}
\end{figure}

We now build on our fiducial model by repeating our sets of runs for various values of $\bsurf$. Figure~\ref{fig:amf-beta} shows an example of the dependence of $\FJ$ on $\bsurf$ for model \texttt{surf}, with $\FJ$ exhibiting fastest damping for $\bsurf\approx1$ in agreement with results by \citet{miranda-rafikov-2020a} and \citet{zhang-zhu-2020}.

\begin{figure}
	\includegraphics[width=\columnwidth]{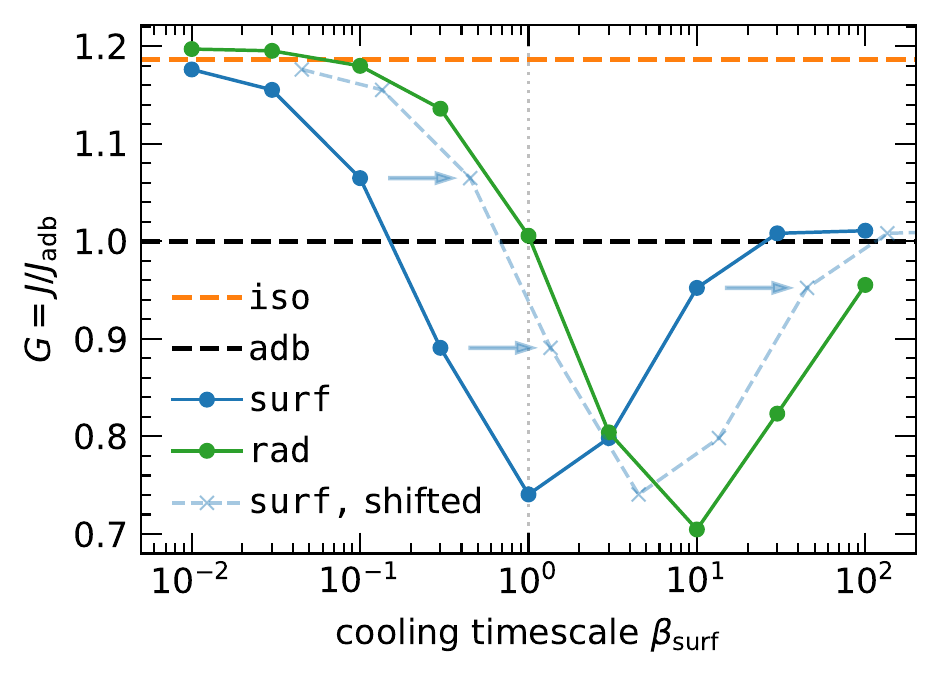}
	\caption{Dependence of the normalized, radially integrated AMF $G$ (Eq.~\eqref{eq:factor-G}) on the surface cooling timescale $\bsurf$ for our models with $\tau=10$ and various treatments of cooling. We note that model \texttt{surf} (blue curve) shows a minimum for $\bsurf\approx1$, which can also be inferred from Fig.~\ref{fig:amf-beta}. Including radiative diffusion (model \texttt{rad}, green) should then simply shift the curve towards larger $\bsurf$ by a factor of $\bsurf/\btot$ (see Eq.~\eqref{eq:factor-f}), which is not exactly the case. Shifting the blue curve by $\bsurf/\btot$ (dashed blue) instead shows that the two curves are still offset horizontally by a factor of $\approx1.5$.}
	\label{fig:G-beta}
\end{figure}

We then compute the factors $G^\text{surf}$ and $G^\text{rad}$ via Eq.~\eqref{eq:factor-G} and plot them as a function of $\bsurf$ in Fig.~\ref{fig:G-beta}. Focusing on the solid blue and green curves, three main takeaways can be extracted from this figure:
\begin{enumerate}
	\item We confirm that radiative damping of planet-driven density waves is strongest and therefore the planet's capacity to open multiple gaps is weakest for $\bsurf\approx1$ in a model where surface emission is the only cooling channel, in agreement with \citet{zhang-zhu-2020}.
	\item We recover both the locally isothermal and adiabatic limits at very small and very large $\bsurf$ respectively, adding to the robustness of our method.
	\item Accounting for both surface cooling and in-plane radiative diffusion (\texttt{rad} models) does not always imply weaker radiative damping of spirals. Instead, the corresponding curve of $G(\bsurf)$ in Fig.~\ref{fig:G-beta} is simply shifted to larger $\bsurf$ compared to model \texttt{surf} but behaves similarly otherwise.
\end{enumerate}

The last point can be explained with the same argument as in the previous section: in-plane radiative diffusion enables an additional cooling channel for spiral shocks and allows them to behave more akin to locally isothermal shocks.

The above also implies that there is a critical cooling timescale $\beta_\text{crit}$ for which the $G$ parameter in \texttt{rad} models shows a minimum, similarly to how $\beta_\text{crit} = \bsurf$ for \texttt{surf} models. Assuming that in this case $\beta_\text{crit}=\btot$, we define the factor $f$ through Eq.~\eqref{eq:factor-f} and then shift the blue curve in Fig.~\ref{fig:G-beta} towards the right by $\bsurf/\btot = f+1 \approx 4.51$. We find that the \texttt{rad} (in green) and shifted \texttt{surf} curves (faint blue), do not quite match, highlighting that the non-local radiative diffusion in our \texttt{rad} models cannot be modeled accurately with a local cooling approach. Instead, the \texttt{surf} models would need to be shifted by a further factor of $\approx 1.5$ in order to overlap with the \texttt{rad} curve.

\begin{figure}
	\includegraphics[width=\columnwidth]{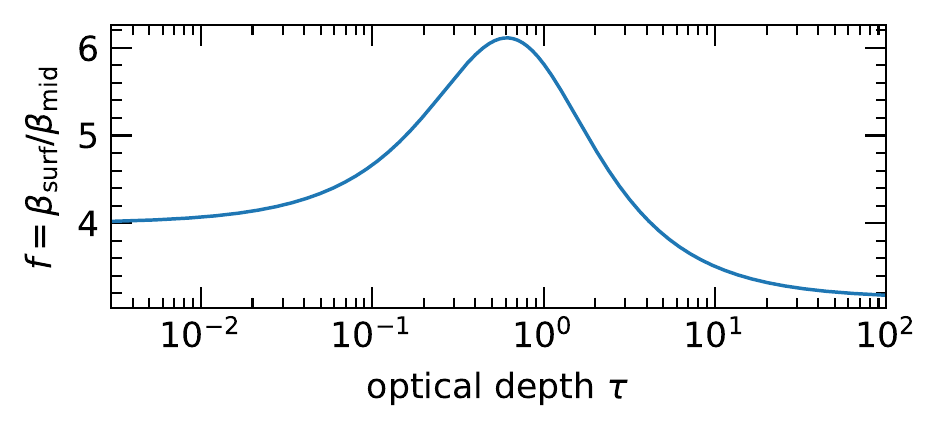}
	\caption{The factor $f$ in Eq.~\eqref{eq:factor-f} as a function of the optical depth $\tau$.}
	\label{fig:f-vs-tau}
\end{figure}

\begin{figure}
	\includegraphics[width=\columnwidth]{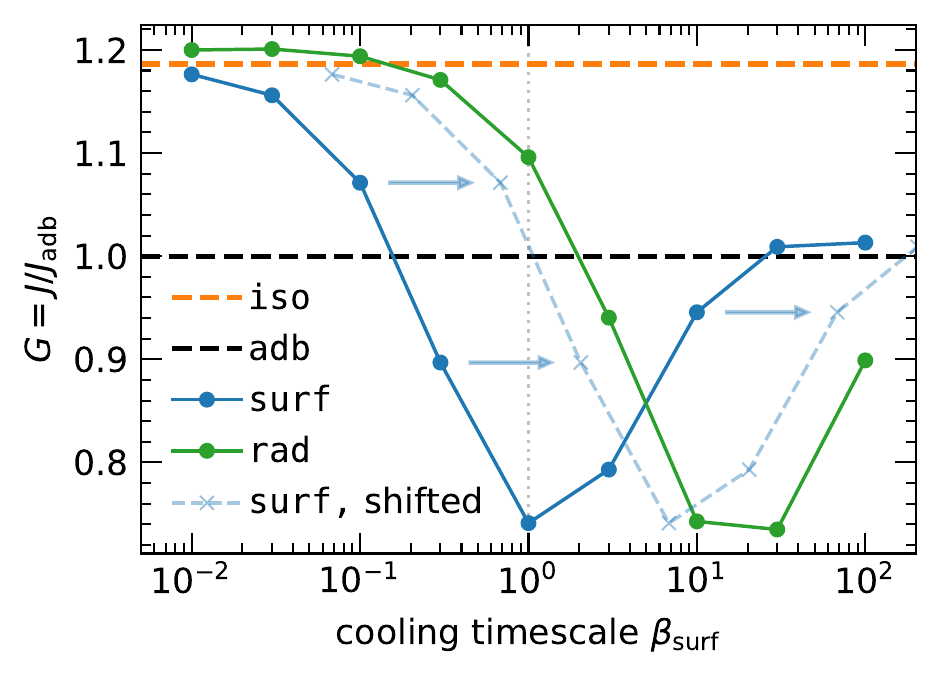}
	\caption{Similar to Fig.~\ref{fig:G-beta} but for models with $\tau=1$, such that the ratio of surface to in-plane cooling is larger. Here, we find that the offset between $G^\text{surf}$ and $G^\text{rad}$ requires a further correction of $\bsurf$ by a factor of $\approx 2$.}
	\label{fig:G-beta-tau1}
\end{figure}

We note that $f$ is a sensitive function of $\tau$, peaking at $\tau\approx0.6$ ($f\approx6.1$, see Fig.~\ref{fig:f-vs-tau}). Essentially, this means that the green curve in Fig.~\ref{fig:G-beta} would shift further to the right (towards longer $\beta$, as in-plane cooling would be more efficient), while the blue curve would not be displaced as in-plane cooling is not considered. To showcase this effect, and to add to the robustness of our results, we repeat our models for $\tau=1$ and present our findings in Fig.~\ref{fig:G-beta-tau1}. Once again the \texttt{rad} and shifted \texttt{surf} models do not quite match, with the two curves being offset by a factor of $\approx2$, but at the qualitative level the agreement between these models is reasonably good.

\subsection{Comparison with various $\beta$-cooling prescriptions}
\label{sub:amf-beta-vs-Q}

Until this point we implemented our heating and cooling prescriptions similar to \citet{ziampras-etal-2020b} by directly adding radiative source terms ($\Qirr$, $\Qcool$, $\Qrad$) to the energy equation. Previous studies on this topic, however, used the simpler and more popular $\beta$-cooling approach by defining the cooling timescale similarly to Eqs.~\eqref{eq:bsurf}~or~\eqref{eq:btot} and then implementing a relaxation term similar to $\Qrelax$ in Eq.~\eqref{eq:Qrelax} in the energy equation \citep[e.g.,][]{miranda-rafikov-2020a,miranda-rafikov-2020b,zhang-zhu-2020}.
While our findings agree qualitatively with these studies in that gap opening is weakest in models with surface cooling and $\bsurf\sim1$ \citep{miranda-rafikov-2020a,zhang-zhu-2020} and that in-plane cooling dramatically changes spiral AMF \citep{miranda-rafikov-2020b}, we would like to quantify the differences between the two approaches. 

To investigate this, we carry out two more sets of models to complement our \texttt{surf} and \texttt{rad} runs. By substituting the balance between $\Qirr$ and $\Qcool$ with a cooling term $\Qrelax$ where $\beta=\bsurf$ (models tagged ``\texttt{bsurf}''), and by further implementing in-plane cooling via $\beta=\btot$ through Eqs.~\eqref{eq:bmidH}~\&~\eqref{eq:btot} in a similar manner (models ``\texttt{btot}''), we aim to compare models in the pairs \texttt{surf}--\texttt{bsurf} and \texttt{rad}--\texttt{btot} with each other. Table~\ref{table:eos} provides a description of all physical options and implementations.

\begin{figure}
	\includegraphics[width=\columnwidth]{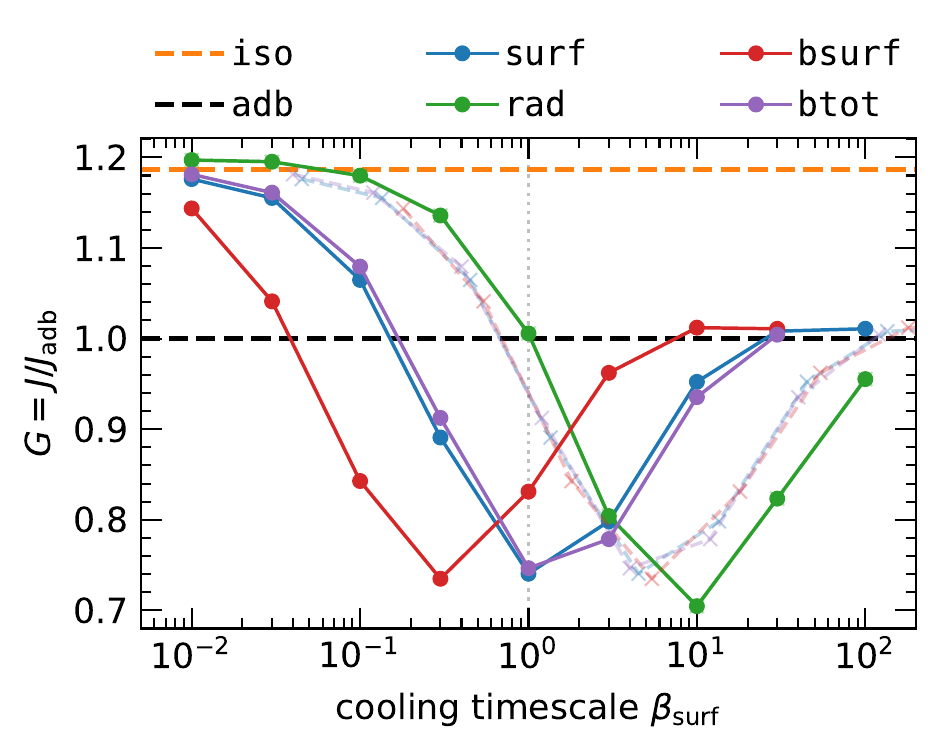}
	\caption{The parameter $G(\bsurf)$ similar to Fig.~\ref{fig:G-beta} ($\tau=10$), this time including models where the $\beta$ cooling approach was used (\texttt{bsurf}, \texttt{btot}). They are shifted towards shorter $\beta$ by a factor of 4 compared to models with explicit source terms (\texttt{surf} and \texttt{rad}, respectively). Accounting for this factor and the ratio $\bsurf/\btot$ (when applicable) shows that all local cooling prescriptions behave similarly (faint curves), matching \texttt{rad} models only qualitatively.
 }
	\label{fig:G-beta-all}
\end{figure}

\begin{figure}
	\includegraphics[width=\columnwidth]{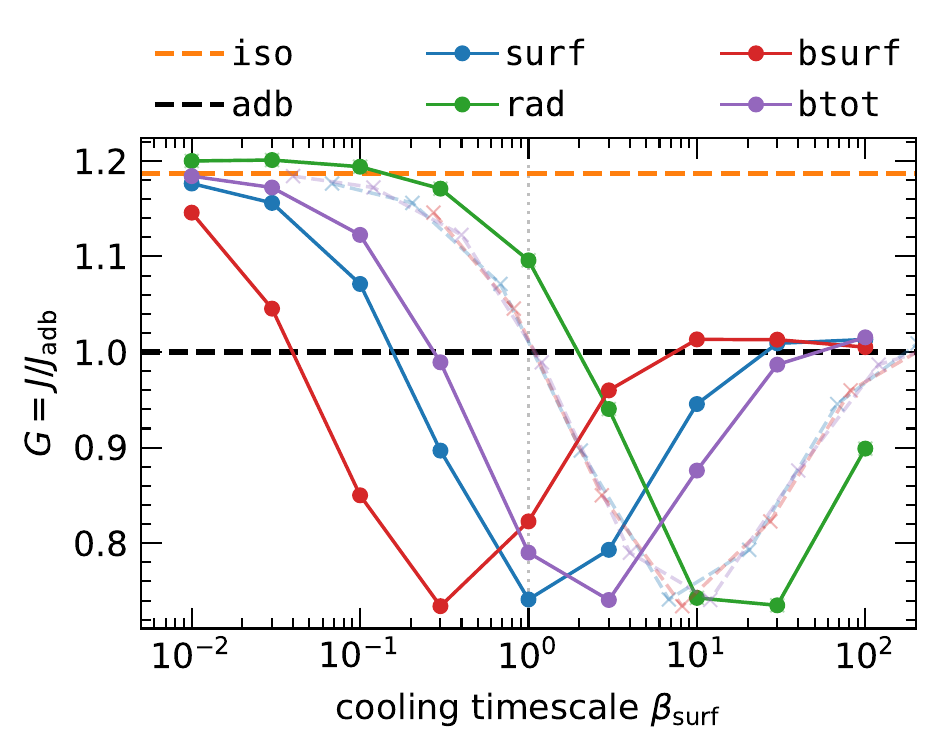}
	\caption{Same as Fig.~\ref{fig:G-beta-all}, for $\tau=1$, showing that our models are robust regardless of $\tau$. Here it becomes clear that the overlap of models \texttt{surf} and \texttt{btot} for $\tau=10$ (blue and purple curves in Fig.~\ref{fig:G-beta-all}) was simply a coincidence.}
	\label{fig:G-beta-all-tau1}
\end{figure}

Figure~\ref{fig:G-beta-all} shows $G(\bsurf)$ for the four approaches described above and for $\tau=10$. The purple curve, which uses Eqs.~\eqref{eq:bsurf}, \eqref{eq:bmidH},~and \eqref{eq:btot}, is offset towards longer $\beta$ when compared to the red curve \citep[prescription of][]{miranda-rafikov-2020a,zhang-zhu-2020} by the factor $1+f$ described in Eq.~\eqref{eq:factor-f}, showing that their findings using a $\beta$-cooling approach are qualitatively consistent with our explicit treatment of heating and cooling terms in terms of spiral AMF. The same behavior can be seen in Fig.~\ref{fig:G-beta-all-tau1}, where $\tau=1$.

However, these figures also show that model pairs that treat the same cooling mechanisms with different implementations (\texttt{surf}--\texttt{bsurf} and \texttt{rad}--\texttt{btot}) do not match each other. We identify that the reason behind this mismatch is inherent to the treatment of cooling using a $\beta$ approach where $\Qrelax\propto\beta^{-1}\propto \Qcool\propto T^4$, which overestimates the true cooling timescale by a factor of 4 when small deviations around an equilibrium state are considered. Our setup reflects such a case, with temperature deviations in the order of $\delta T/T_0 \lesssim 0.1$ around spiral arms for $\Mp=0.3\,\Mth$ (see top left panel in Fig.~\ref{fig:fiducial-planet-tmp}). This factor of 4 has also been reported by \citet{dullemond-etal-2022}. We provide a detailed derivation of this factor in Appendix~\ref{apdx:Q-vs-beta}.

Shifting $\beta$ models to the right by this factor of 4 and further adjusting models \texttt{surf} and \texttt{bsurf} by the factor $1+f$ as in Sect.~\ref{sub:amf-beta} (faint curves in Figs.~\ref{fig:G-beta-all}~\&~\ref{fig:G-beta-all-tau1}) shows that, once again, no prescription matches the results with FLD (\texttt{rad}) at the quantitative level. This highlights the importance of modeling FLD self-consistently instead of using a local approximation.

We note that the overlap between models \texttt{surf} (blue) and \texttt{btot} (purple) in Fig.~\ref{fig:G-beta-all} before shifting is simply a coincidence. Model \texttt{surf} overestimates the true cooling timescale $\btot$ by a factor of $1+f$ due to the lack of in-plane cooling (see Eq.~\eqref{eq:factor-f}), while model \texttt{btot} overestimates $\btot$ by a factor of 4 due to using $\beta$ cooling. Since $1+f\approx 4.5$ for $\tau=10$, the two curves seemingly overlap.
For $\tau=1$ (Fig.~\ref{fig:G-beta-all-tau1}) we have $1+f\approx6.8$, and the two curves are distinctly different.

To test the limits of our approach in the nonlinear regime, we repeated a series of models with $\tau=10$ and $\Mp=3\,\Mth$. Our results can be found in Appendix~\ref{apdx:nonlinear}, where we show that the quoted factor of 4 reduces to approximately 3.5 for this more massive planet. To further check whether our choice of $\hp$ could affect our results, we carried out a set of models with $\hp=0.1$ and a resolution of 32 cells per scale height ($N_R\times N_\phi=600\times2048$) and found very good agreement with our models with $\hp=0.05$. These results are shown in Appendix~\ref{apdx:aspect-ratio}. 

\FloatBarrier

\subsection{An approximate recipe for FLD}
\label{sub:approx-fld}

\begin{figure}
	\includegraphics[width=\columnwidth]{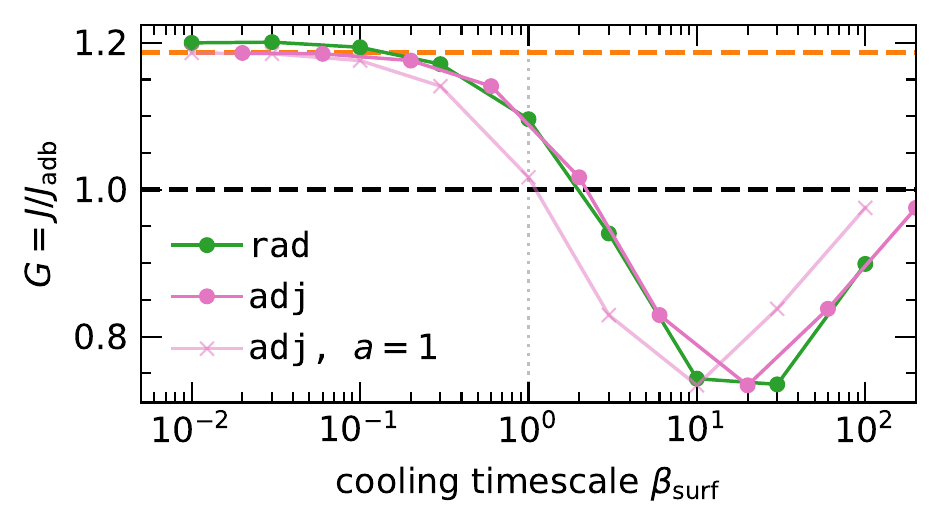}
	\caption{Similar to Fig.~\ref{fig:G-beta-all-tau1} ($\tau=1$), but comparing models with explicit treatment of FLD (green) to the approximate recipe given by Eq.~\eqref{eq:badj} that accounts for in-plane cooling through $\beta$ (pink). We find an excellent agreement between the two curves.}
	\label{fig:G-beta-adj}
\end{figure}

Based on the above, one could in principle account for all necessary factors that differentiate the $\beta$ models using Eqs.~\eqref{eq:bmidH}~\&~\eqref{eq:btot} and our fully radiative FLD models and arrive at a model that behaves similar to model \texttt{rad} in terms of AMF. To test this, we design an adjusted cooling timescale that aims to account for the factor 4 discussed above (see Appendix~\ref{apdx:Q-vs-beta}) and write:
\begin{equation}
	\label{eq:badj}
	\beta_\text{adj} = \frac{1}{4a}\,\btot =\frac{1+f}{4a}\,\bsurf,\qquad a\approx1.5\text{--}2,
\end{equation}
where $\bsurf$ and $f$ are defined through Eqs.~\eqref{eq:bsurf}~\&~\eqref{eq:factor-f}, respectively, and $a$ is a correction factor motivated by the offset between \texttt{surf} and \texttt{rad} models in Figs.~\ref{fig:G-beta-all}~\&~\ref{fig:G-beta-all-tau1}. A comparison between a set of models using this recipe---tagged ``\texttt{adj}''---and our \texttt{rad} models in terms of $G(\bsurf)$ is shown in Fig.~\ref{fig:G-beta-adj} for $\tau=1$, and we find excellent agreement between the two sets of models as far as this metric is concerned for $a=2$.

\begin{figure}
	\includegraphics[width=\columnwidth]{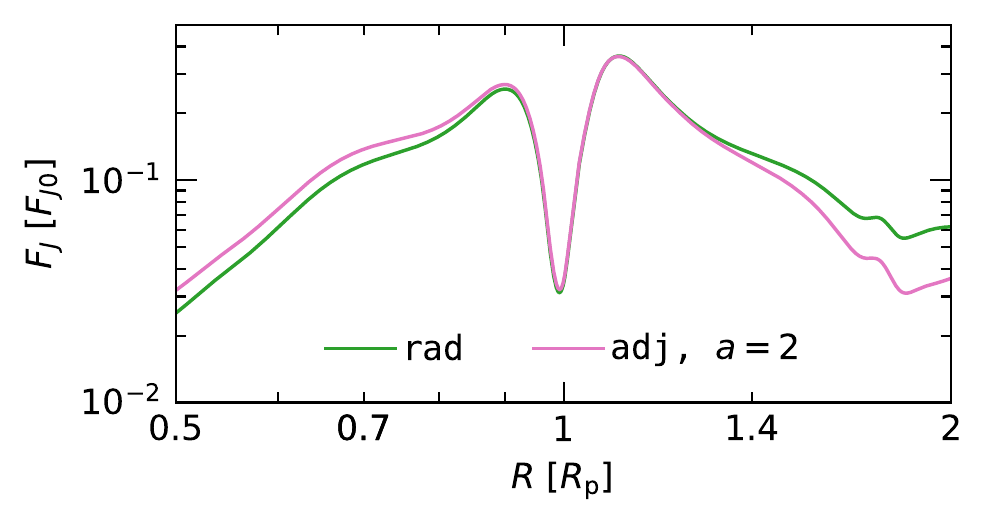}
	\caption{Radial profiles of $\FJ$ for models \texttt{rad} (green) and \texttt{adj} (pink) for $\tau=1$, $\bsurf=1$ and $a=2$. While the two models have the same $G$ (see Fig.~\ref{fig:G-beta-adj}), their $\FJ$ profiles are somewhat different, mainly far from the planet.}
	\label{fig:amf-adj}
\end{figure}

However, while $G$ is the same between the two sets of models, that does not mean that the radial AMF profiles $\FJ(R)$ will also match, as seen in Fig.~\ref{fig:amf-adj} for $\tau=1$, $\bsurf=1$ and $a=2$. We find that $\FJ$ is higher (lower) in the inner (outer) disk for model \texttt{adj} compared to model \texttt{rad}, even though integrating $\FJ$ radially yields the same result.

Furthermore, the two sets of models are quite different in terms of the structure of the spiral arms themselves. As we discussed in Sect.~\ref{sec:physics-motivation} and highlighted in Fig.~\ref{fig:example-beta}, our FLD models allow the diffusion of heat into the pre-shock region, something that $\beta$ models fail to capture.

\begin{figure}
	\includegraphics[width=\columnwidth]{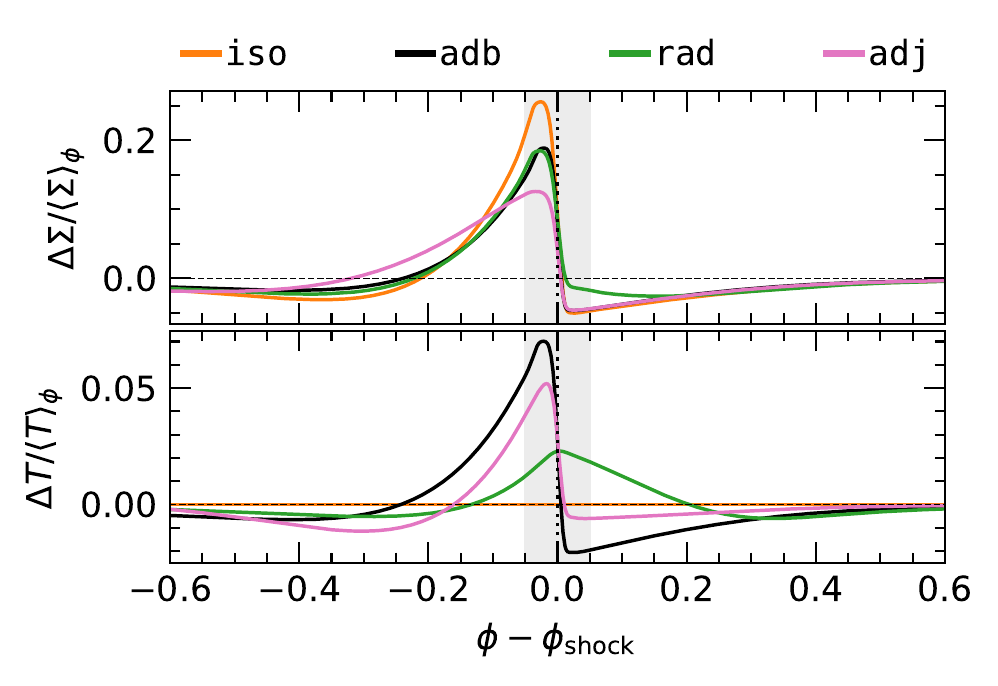}
	\caption{A slice along the spiral shock front at $R=1.2\,\Rp$, showing its azimuthal structure. While models \texttt{rad} and its approximate equivalent \texttt{adj} (green and pink curves, respectively) are functionally identical in terms of our AMF metric $G$ (see Fig.~\ref{fig:G-beta-adj}), the shock structure is completely different. Thanks to in-plane radiative diffusion, model \texttt{rad} shows a weaker temperature spike and thus behaves more similar to the locally isothermal model (\texttt{iso}, orange curves). See also Fig.~\ref{fig:shock-front}.}
	\label{fig:rad-corr-shock}
\end{figure}

This effect is illustrated in Fig.~\ref{fig:rad-corr-shock}, where we compare the surface density and temperature structure of the shock front in a model with $\bsurf=3$, $\tau=1$ for these two implementations of in-plane cooling using an azimuthal slice at $R=1.2\,\Rp$. Visible is a stark contrast in temperature between models \texttt{rad} (green) and \texttt{adj} (pink), with the former showing a heat excess in the pre-shock region (right half) while the latter shows a clear, sharp temperature spike along the shock front.

\subsection{Section summary}

In summary, we confirm the findings of \citetalias{miranda-rafikov-2020b} in that in-plane cooling strongly affects the spiral wave AMF, and by extension the propensity for planet-induced gap opening. We further show that fully radiative models (\texttt{rad}) are distinctly different from \texttt{surf} models, as simply accounting for the ratio $\bsurf/\btot$ locally does not yield a good match between the two models, with an offset of $\approx1.5$--2 along the $\beta$ axis depending on $\tau$. Nevertheless, the existence of a minimum in $G$ naturally splits the $\beta$ space into a fast- and slow-cooling regime, where omitting the effects of in-plane cooling under- and overestimates the planet's capacity to open multiple gaps, respectively. We investigate this further in Sect.~\ref{sec:real-systems}.  

We also find that a local, $\beta$ prescription with $\beta=\bsurf$, similar to that by \citet{miranda-rafikov-2020a} and \citet{zhang-zhu-2020}, results in an effective cooling timescale that is larger by a factor of 4 than in our \texttt{surf} models due to their use of $\beta$ cooling with a term $\Qcool\propto T^4$. While this does not take away from their findings, as they simulate a generic system to showcase the effect of in-plane cooling, this difference becomes important for ALMA disks where the cooling timescale in the observable range is commonly $\bsurf\sim0.1\text{--}10$.

To reconcile this, we design a recipe for $\beta$ cooling (Sect.~\ref{sub:approx-fld}) that accounts for all theoretically motivated differences between $\beta$ models and our model \texttt{rad} which treats radiation transport directly. While reproducing the AMF behavior of the density waves reasonably well, however, this adjusted $\beta$ approach misses the diffusion of heat around spiral shocks, something that affects their density and temperature structure. In the next section, we will demonstrate how this adjusted $\beta$ recipe performs at predicting gap profiles carved by planets in realistic disks with radially-varying quantities. 

\section{Application to real systems}
\label{sec:real-systems}

In the previous section we explored the differences in AMF between models that omit or treat in-plane radiation transport, and used the parameter $G$ as a proxy for estimating the planet's gap-opening capabilities. We found two regimes of $\bsurf\gtrsim1$ and $\bsurf\lesssim1$, where the omission of in-plane cooling under- and overestimates spiral radiative damping respectively. Since stronger radiative damping translates to a single, deep gap rather than several, shallow gaps \citepalias{miranda-rafikov-2020b}, accounting for in-plane cooling will result in fewer gaps for $\bsurf\gtrsim1$ and vice versa for $\bsurf\lesssim1$.

Since the latter scenario is more common at radii where planets are predicted to exist in ALMA-observed systems ($R\sim50$\,au), accounting for in-plane radiative diffusion should typically result in more rings and gaps than when neglecting it. This is consistent with the findings of \citetalias{miranda-rafikov-2020b}, supports the hypothesis that a planet can be responsible for multiple gaps \citep[e.g.,][]{zhang-etal-2018}, and explains the weaker substructure in \citet{ziampras-etal-2020b}, where the effect of in-plane cooling was missed. The opposite is true, however, in the slower-cooling, more optically thick inner disk ($\bsurf\gtrsim10$, $\tau\gtrsim10$). Here, in-plane radiative diffusion serves to reduce the AMF of planetary spirals close to the planet, reducing their ability to open multiple gaps.

In this section, we test these assumptions by running long-term models of systems with parameters inspired by ALMA observations.

\subsection{Estimates of $\bsurf$ and $\tau$ in ALMA systems}
\label{sub:alma-estimates}

As a proof of concept, we compute $\tau$ and $\bsurf$ at the position of annular features (rings, gaps) for several systems present in the DSHARP survey \citep{andrews-etal-2018}. We perform this by combining the list of features in Table~1 of \citet{huang-etal-2018} with estimates of the surface density $\Sigma$ at each position using Table~3 in \citet{zhang-etal-2018} for all applicable systems. Stellar properties ($\Mstar$, $\Lstar$) are given by \citet{andrews-etal-2018}. We then assume for a background state a passively heated disk model where $\Qirr = -\Qcool$, or
\begin{equation}
	T = T_0 R^{-\nicefrac{1}{2}},\qquad T_0 = \left(\frac{\theta\Lstar}{8\pi\sigmaSB}\right)^{\nicefrac{1}{4}},
\end{equation}
and compute the Rosseland and Planck mean opacities using the joint power-law formula of \citet{lin-papaloizou-1985} \citep[for details see Table~1 in][]{mueller-kley-2012}, which for our models practically reduces to $\kappaR=\kappaP=2\times10^{-4}\, (T/\text{K})^2$ $\text{cm}^2$ per gram of gas. We finally calculate the optical depth $\tau$ and the cooling timescale $\bsurf$ using Eqs.~\eqref{eq:Qcool}~\&~\eqref{eq:bsurf}.

\begin{figure}
	\includegraphics[width=\columnwidth]{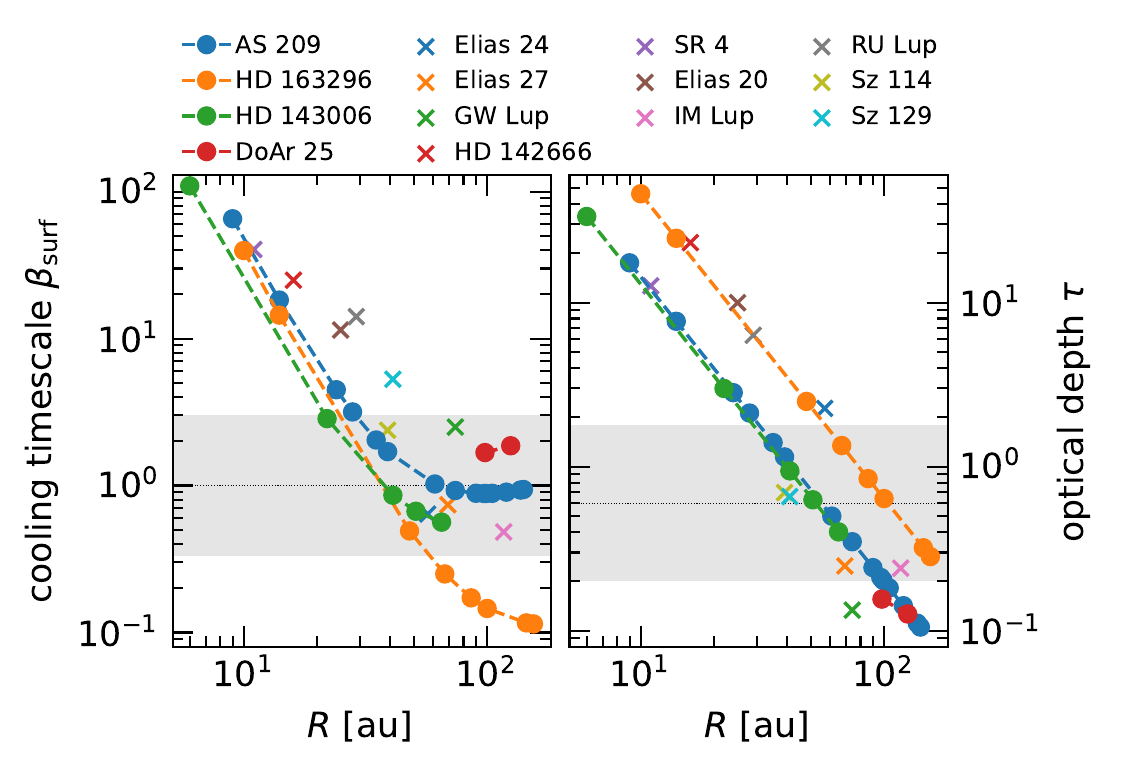}
	\caption{Surface cooling timescale $\bsurf$ and optical depth $\tau$ for a variety of DSHARP systems \citep{andrews-etal-2018}, using estimates of gas and dust properties discussed in Sect.~\ref{sub:alma-estimates}. Shaded bands mark areas where the inclusion of in-plane cooling can strongly affect gas thermodynamics and therefore planet-driven gap opening.}
	\label{fig:dsharp-beta-tau}
\end{figure}

Our results are shown in Fig.~\ref{fig:dsharp-beta-tau}, where we mark interesting regions---where treating in-plane cooling is expected to produce significant differences to surface-cooling-only models---with shaded bands. Notably, we find a large cluster of features for $\bsurf\sim 1$ and $\tau\sim0.6$, the latter being consistent with estimates of the optical depth by \citet{dullemond-etal-2018}. We note, however, that given our liberal estimates of $\Sigma$ and our relatively simple opacity model, our estimated values of $\bsurf$ and $\tau$ might not be very accurate.

Based on these estimates, we choose the systems AS~209 and Elias~20 to represent the marginally optically thin ($\bsurf\sim1$, $\tau\lesssim1$) and optically thick regimes ($\bsurf\sim10$, $\tau\sim10$). For these two systems the inclusion of in-plane cooling should result in an enhancement/suppression of gap opening compared to models with surface cooling only, respectively. We describe our setup for these simulations in the next section.

\subsection{Model setup}
\label{sub:model-setup}

Our setup is very similar to that described in Sect.~\ref{sub:amf-model} with minor changes. Namely, we base our initial conditions on our estimates in Sect.~\ref{sub:alma-estimates} and choose initial surface density and temperature profiles $\Sigma_0(R) \propto R^{-1}$ and $T(R)\propto R^{-1/2}$, as well as reference aspect ratios $\hp\approx0.08$ for the two systems, the latter given by a balance between $\Qirr$ and $\Qcool$. The Rosseland and Planck mean opacities now follow the recipe of \citet{lin-papaloizou-1985} as well. Our initial conditions are consolidated in Table~\ref{table:alma-init} and plotted on the left panels of Fig.~\ref{fig:systems-gas}.
\begin{table}
	\begin{center}	
	\caption{Parameters used in our models of AS~209 and Elias~20. Values with a subscript `p' refer to the planet's position.}
	\label{table:alma-init}
	\begin{tabular}{l | c | c}
		\hline
		Quantity & AS~209 & Elias~20 \\
		\hline
		$\Mstar$ [$\Msun$] & 0.83 & 0.48\\
		$\Lstar$ [$\Lsun$] & 1.41 & 2.24\\
		$\Rp$ [au] & 99 & 29 \\
		$\Mp$ [$\Mjup$] & 0.144 & 0.091 \\
		$\Sigma_\text{p}$ [g/cm${}^2$] & 10 & 80\\
		$T_\text{p}$ [K] & 14.2 & 29.7\\
		$\hp$ & 0.082 & 0.084\\
	\end{tabular}
	\end{center}	
\end{table}

Similar to Sect.~\ref{sub:amf-model}, we use grids logarithmically spaced in $R\in[0.1,4]\,\Rp$ and with a resolution of 16 cells per scale height at $\Rp$, for a grid size of $N_R \times N_\phi = 721 \times 1225$ for AS~209 and $701\times1189$ for Elias~20. This resolution strikes a good balance between appropriately resolving gas dynamics, maintaining feasible computation runtimes, and comparing with the literature.

The planets have a mass of $\Mp=0.3\,\Mth$ to be consistent with our analysis in the previous section. This translates to $\Mp=0.144\,\Mjup$ and $0.09\,\Mjup$ for AS~209 and Elias~20, respectively, which are reasonable values compared to the fits of \citet{zhang-etal-2018} if we extrapolate to the lower viscosity of $\alpha=10^{-5}$ that we use. The planets are then embedded at $\Rp=99$\,au and 29\,au for AS~209 and Elias~20, respectively. We note that our goal here is not to reproduce these two systems optimally, but instead to showcase the importance of in-plane cooling in gap opening in the optically thin and thick regimes, and therefore the exact values of $\Mp$ and $\alpha$ are not important.

Both systems are then evolved for 500 planetary orbits using the following four cooling models: locally isothermal (``\texttt{iso}''), surface cooling (``\texttt{surf}''), surface and in-plane cooling (``\texttt{rad}''), and adjusted $\beta$ cooling (``\texttt{adj}''). The first three are described in Table~\ref{table:eos}, and the latter is introduced in Eq.~\eqref{eq:badj} and discussed in Sect.~\ref{sub:approx-fld}.

\subsection{Results}
\label{sub:dsharp-results}

In this section we present our findings for the models of AS~209 and Elias~20 that were described above. We first discuss differences in the gas profiles, and continue with simplistic synthetic observations of continuum emission.

\subsubsection{Gas profiles}
\label{sub:dsharp-gas}

\begin{figure*}
	\includegraphics[width=\textwidth]{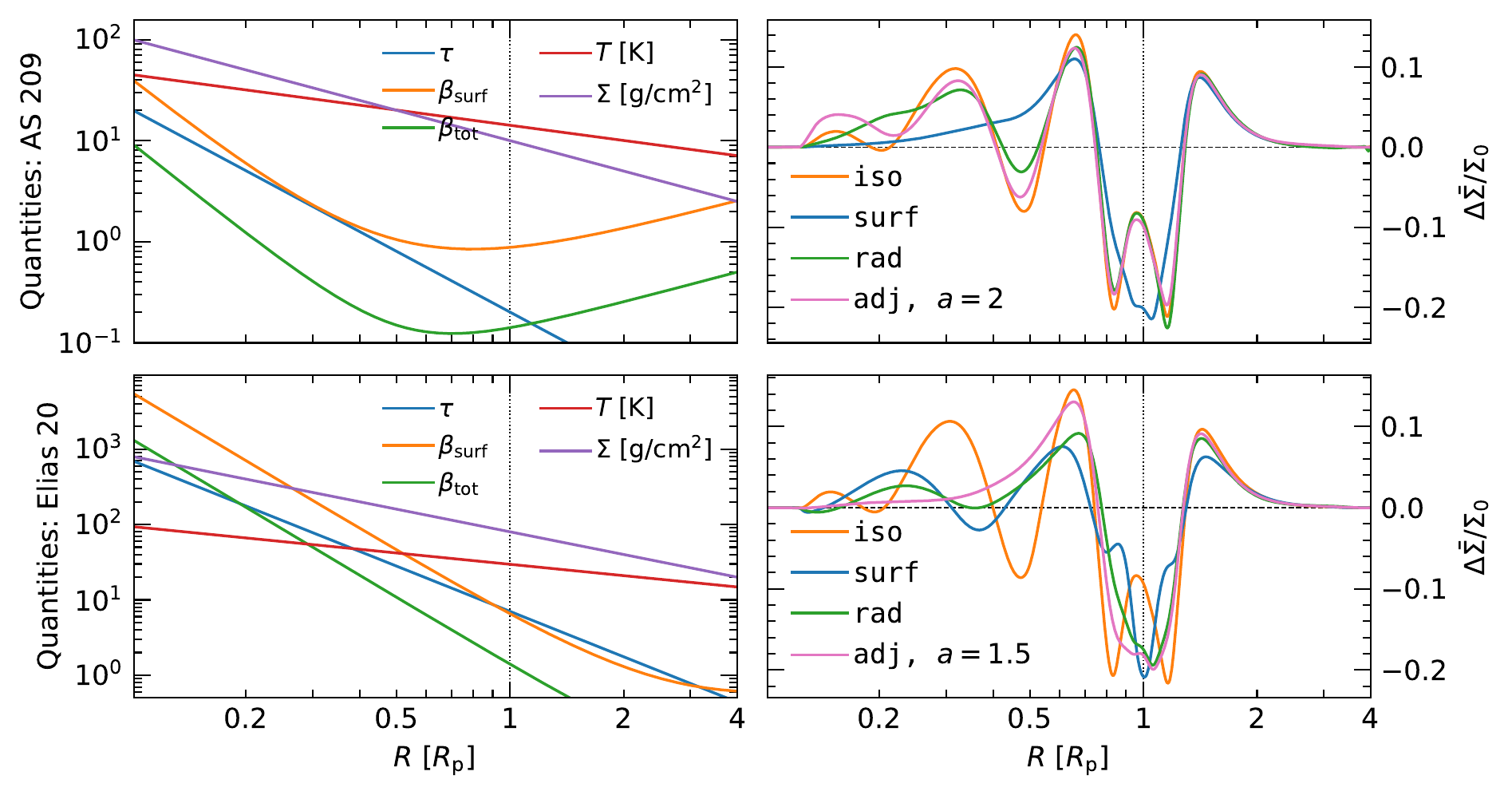}
	\caption{Gas profiles for AS~209 (top) and Elias~20 (bottom). Left panels show various relevant quantities at $t=0$, while right panels show azimuthally averaged surface density deviations after 500 planetary orbits. Undulations suggest possible rings or gaps, which are explored further in Sect.~\ref{sub:dsharp-dust}. We note a moderate damping of such undulations for models with cooling when compared to the locally isothermal one (\texttt{iso}, orange curves on the right), and the different behavior of models with and without in-plane cooling (\texttt{rad}, green and \texttt{surf}, blue, respectively). Treating in-plane cooling with our adjusted $\beta$ recipe via Eq.~\eqref{eq:badj} (model \texttt{adj}, pink) matches the fully radiative model \texttt{rad} reasonably well for AS~209, but not so much for Elias~20.}
	\label{fig:systems-gas}
\end{figure*}

The gas surface density structure after 500 orbits is shown in the right panels of Fig.~\ref{fig:systems-gas}. We note the lack of a secondary gap at $R\approx0.5\,\Rp$ as well as a deeper primary gap for model \texttt{surf} for AS~209 (panel \emph{b}), compared to other models, fully consistent with the findings of \citet{ziampras-etal-2020b}. At the same time, the two models that treat in-plane cooling (\texttt{rad} and \texttt{adj}), while slightly different further into the inner disk, both capture the secondary gap as well as the bump inside the planet's corotating region, leading to the possibility of two additional rings compared to model \texttt{surf}. In other words, we confirm that in the (marginally) optically thin regime ($\tau\lesssim1$, $\bsurf\sim1$) the omission of in-plane cooling incorrectly weakens the planet's capacity to open multiple gaps.

Regarding Elias~20 (panel \emph{d} of Fig.~\ref{fig:systems-gas}), comparing models \texttt{surf} (blue) and \texttt{rad} (green) indicates that in-plane cooling slightly suppresses the opening of a secondary gap at $R\approx0.4\,\Rp$ and results in weaker radial features inside the planet's corotating region. This is once again in agreement with our predictions in Sect.~\ref{sec:amf-study}, validating our findings.

However, we find that model \texttt{adj}, which is designed to approximately mimic the effects of in-plane radiative diffusion, results in a rather different radial structure for Elias 20. This is particularly obvious from the lack of a secondary undulation in the inner disk. This is not entirely unexpected: while models \texttt{adj} and \texttt{rad} are functionally identical in terms of the parameter $G$ (a proxy for the AMF), that does not mean their AMF profiles $\FJ(R)$ will also be identical (see Fig.~\ref{fig:amf-adj}). In addition, we showed in Sect.~\ref{sub:approx-fld} that there are significant differences in the density and temperature structure around shock fronts between the two models (see Fig.~\ref{fig:rad-corr-shock}). These effects, in conjunction with our temperature-dependent opacity model \citep[see also Appendix~\ref{apdx:Q-vs-beta} and][]{dullemond-etal-2022} and the existence of radial gradients in $\Sigma$, $\kappa$, $\tau$, and $\beta$---which did not exist in our simple models in Sect.~\ref{sec:amf-study}---shows that this adjusted $\beta$ approach might not always be applicable for realistic disks.

\subsubsection{Continuum emission}
\label{sub:dsharp-dust}

In the interest of comparing our findings with observations, we follow up with synthetic observations of millimeter emission similar to \citetalias{miranda-rafikov-2020b}. This is achieved by postprocessing our azimuthally-averaged disk profiles with a simple model of radial dust drift and computing the resulting emission flux at a wavelength of 1\,mm. Our method is described in detail in Appendix~\ref{apdx:dust}.

\begin{figure}
	\includegraphics[width=\columnwidth]{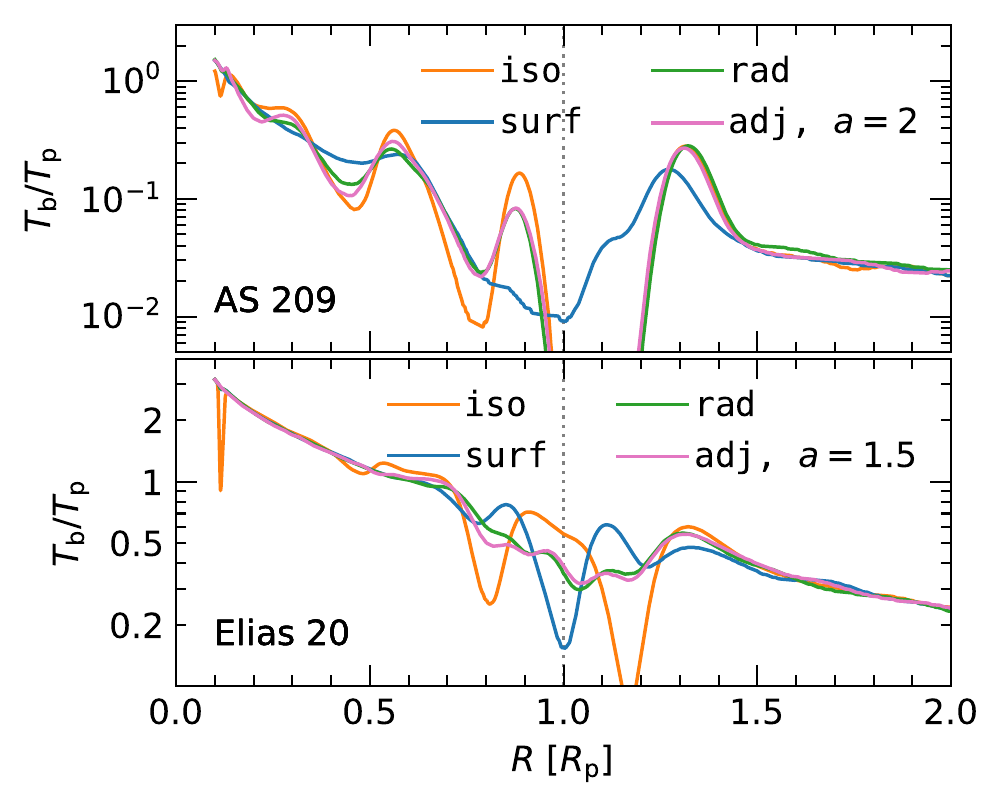}
	\caption{Radial profiles of brightness temperature $\Tb$ for mm-sized dust grains, normalized to the temperature at the planet's location $\Tp$. Model \texttt{surf} shows distinctly different structure, with too few and too many rings for AS~209 (top) and Elias~20 (bottom), respectively. On the other hand, models \texttt{rad} and \texttt{adj} agree quite well.}
	\label{fig:dust-1D}
\end{figure}

\begin{figure}
	\includegraphics[width=\columnwidth]{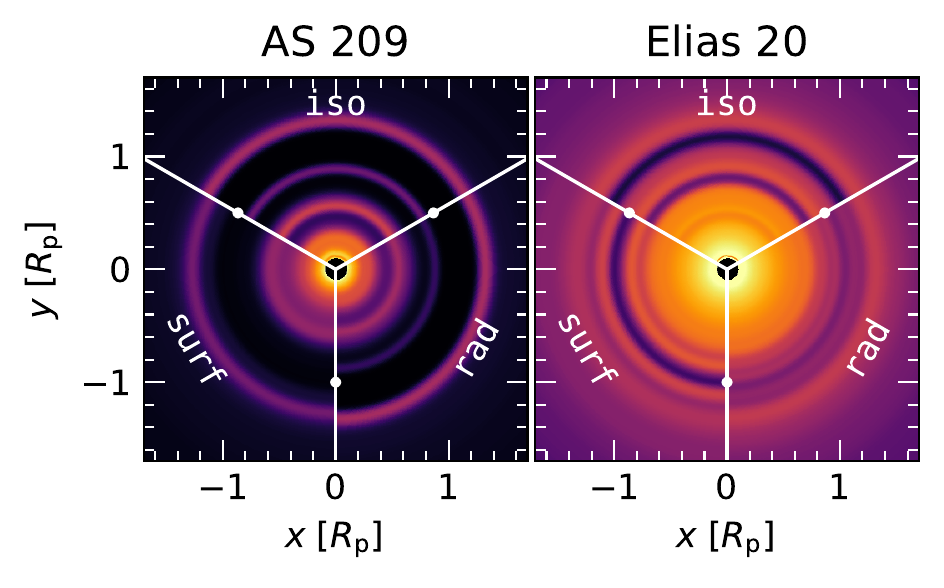}
	\caption{Two-dimensional heatmaps of $\Tb$ similar to Fig.~\ref{fig:dust-1D} to help visualize the differences between models. The planet's radial position is labeled with white dots. The colormap limits match Figs.~\ref{fig:as209-dust}~and~\ref{fig:elias20-dust} for AS~209 and Elias~20, respectively.}
	\label{fig:dust-combined}
\end{figure}

\begin{figure}
	\vcenteredhbox{\includegraphics[width=0.45\columnwidth]{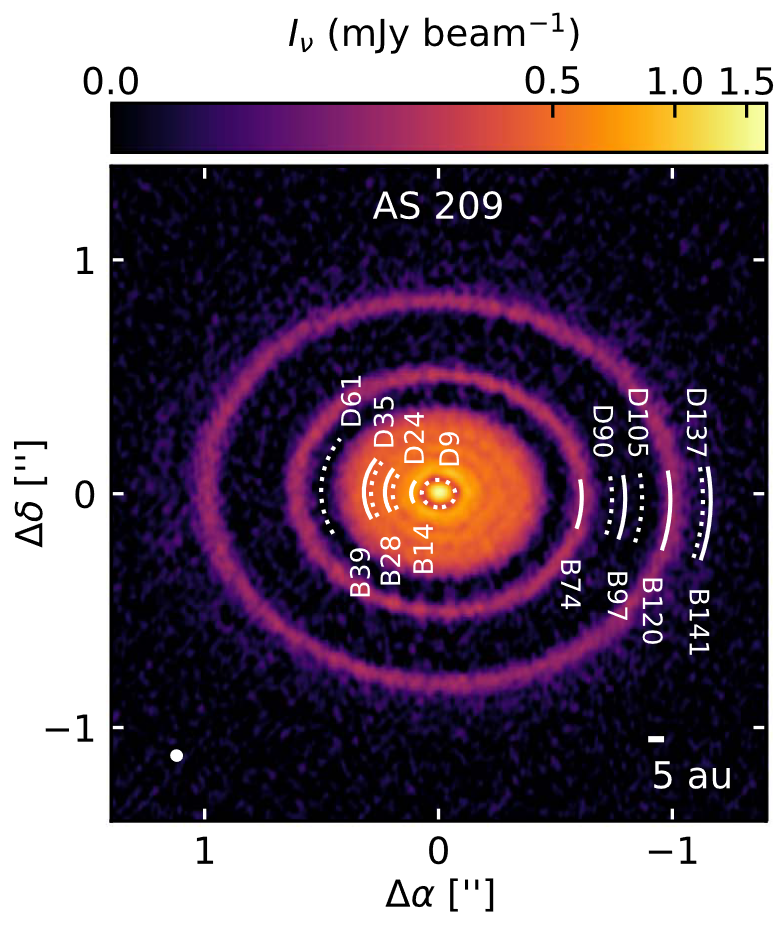}}
	\vcenteredhbox{\includegraphics[width=0.55\columnwidth]{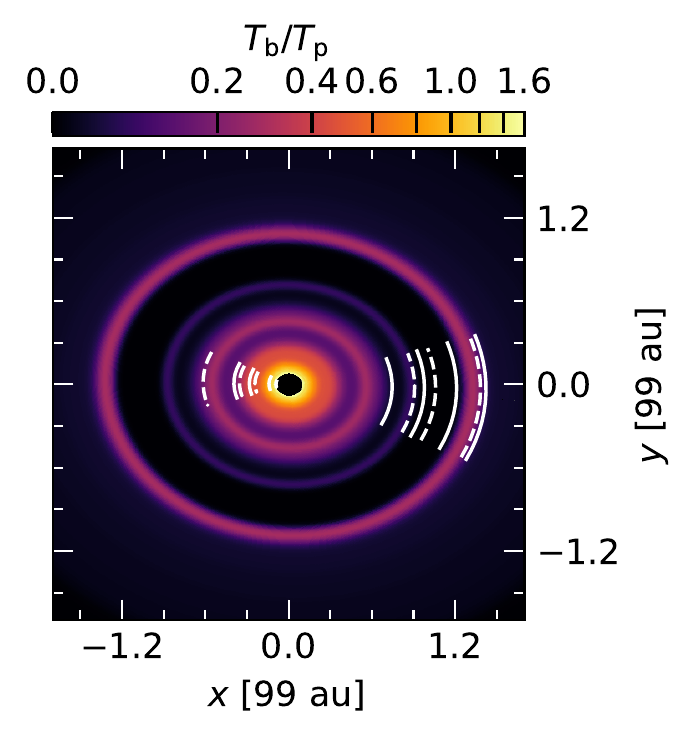}}
	\caption{Left: ALMA observation of AS~209 \citep{huang-etal-2018}. Right: Brightness temperature heatmap for AS~209 for our fully radiative model (``\texttt{rad}'') similar to the left panel of Fig.~\ref{fig:dust-combined} but projected in a way similar to the observation shown in \citet{huang-etal-2018}. Overlaid in white curves are the positions of rings and gaps as reported by that study.}
	\label{fig:as209-dust}
\end{figure}

Our calculated brightness temperature $\Tb$ for AS~209 is shown in the top panel of Fig.~\ref{fig:dust-1D}. We find that, as expected, the model with surface cooling only (``\texttt{surf}'', blue curve) shows distinctly different results. All other approaches reproduce the ring inside the planet's gap region at $R\approx 0.9\,\Rp\approx90$\,au as well as a darker gap at $R\approx0.4\,\Rp\approx40$~au, consistent with the continuum observation \citep{huang-etal-2018} and previous locally isothermal models of this system \citep[e.g.,][]{zhang-etal-2018,ziampras-etal-2020b}. To further demonstrate the differences between models, we provide heatmaps of $\Tb$ in Fig.~\ref{fig:dust-combined} (left panel) as well as a comparison to the ALMA observation of this system in Fig.~\ref{fig:as209-dust}.

\begin{figure}
	\vcenteredhbox{\includegraphics[width=0.46\columnwidth]{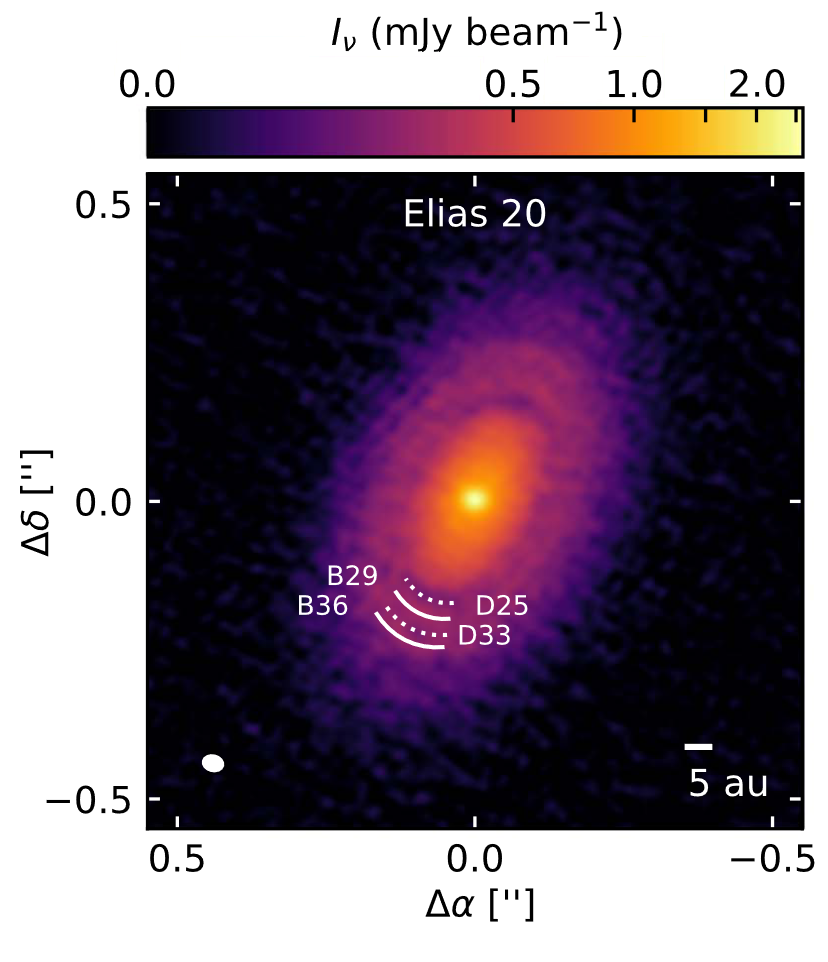}}
	\vcenteredhbox{\includegraphics[width=0.54\columnwidth]{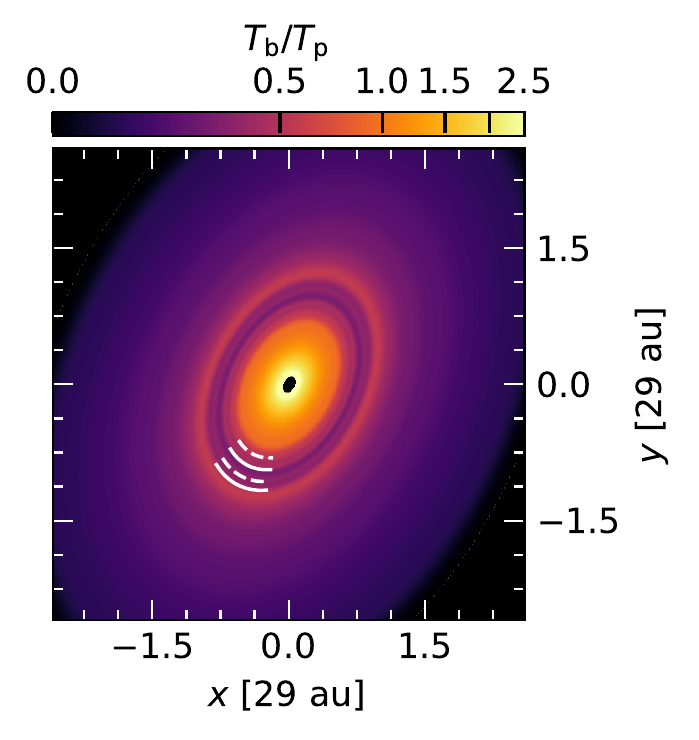}}
	\caption{Left: ALMA observation of Elias~20 \citep{huang-etal-2018}. Right: Brightness temperature heatmap for our \texttt{rad} model of Elias~20, similar to Fig.~\ref{fig:as209-dust} and based on the right panel of Fig.~\ref{fig:dust-combined}.}
	\label{fig:elias20-dust}
\end{figure}

Regarding Elias~20, our results are shown in the bottom panel of Fig.~\ref{fig:dust-1D}, the right panel of Fig.~\ref{fig:dust-combined}, and in Fig.~\ref{fig:elias20-dust}. Here, the \texttt{surf} model shows a clear, deep gap about the planet's orbit (29\,au). Models with a treatment of in-plane cooling (\texttt{rad}, \texttt{adj}), however, show much weaker radial structure. To an extent, this is in line with the observation of Elias~20, which only shows two very shallow gaps about 29\,au \citep{huang-etal-2018}, but a more accurate model is necessary to correctly fit this system. Such a fit is not within the scope of this work, but will be addressed in followup work.

For both systems, we note the very good agreement between the fully radiative model \texttt{rad} and its approximate approach (\texttt{adj}) in the figures described above.


\subsection{Section summary}

In Sect.~\ref{sec:amf-study} we identified two regimes of $\bsurf\gtrsim1$ and $\bsurf\lesssim1$, and showed that in-plane cooling corrects the AMF of planet-driven spirals by damping or enhancing it compared to models with surface cooling only, depending on the regime. We tested this assumption by selecting two representative systems from the DSHARP survey \citep{andrews-etal-2018}, one for each of the two regimes. 

Our findings match our theoretical expectations and previous work very well, in that our \texttt{rad} models consistently recover a more realistic image of both systems regardless of cooling regime when compared to \texttt{surf} models. In addition, we found that model \texttt{adj} recovers a dust structure very similar to that of model \texttt{rad} for both systems, but with differences in the inner disk that could become significant if a more realistic treatment of dust is considered.


\section{Discussion}
\label{sec:discussion}

In this section we touch on our assumptions and their limitations. We also discuss possible applications of our findings.

\subsection{On the quality of our models of real sources}

In Sect.~\ref{sub:dsharp-results} we focused on two ALMA systems, AS~209 and Elias~20, which were part of the DSHARP survey \citep{andrews-etal-2018}. There, we made several assumptions about the properties of the gas and dust in both systems (as well as their ages through our model runtime) and computed millimeter emission maps based on the results of our hydrodynamics models in Sect.~\ref{sub:dsharp-dust}.

It is clear that, looking at Figs.~\ref{fig:as209-dust}~\&~\ref{fig:elias20-dust}, our models do not reproduce the two systems very well: radial features are sometimes observed at different radii, while a number of rings and gaps are missing altogether. We stress that an accurate model of either system is not the goal of this work, but instead the idea that cooling---and in particular in-plane radiation transport---can strongly influence gas hydrodynamics and affect a planet's gap opening capacity. After all, we treat dust transport with a post-processed, radial drift approach, without accounting for the real-time effects of dust concentration and grain size distribution on gas opacity and dynamics.

Nevertheless, our results show that the inclusion of in-plane cooling is a step in the right direction. Accounting for in-plane cooling recovers an additional ring for AS~209 compared to a model that does not, and forms several, shallow gaps about the planet's orbit instead of a single deep gap at $\Rp$ for Elias~20.

\subsection{Using our findings to constrain observables}

\citet{zhang-etal-2018} showed that at least five of the observed rings in AS~209 could be explained by a single planet at 99\,au. While their approach certainly involved some fine-tuning in their parameters in order to match the observed emission ($\hp=0.05$ and $\alpha=3\times 10^{-4}$ at 99\,au), their findings clearly showed that the ``single planet, multiple rings'' scenario is certainly feasible and a realistic fit to observations is possible with the right choice of physical parameters.

This parameter space, however, is not only difficult to explore with high-resolution simulations, but also degenerate to an extent: a disk with low enough $\hp$--$\alpha$ or a massive enough planet can overcome the gap opening threshold \citep{crida-etal-2006} and lead to one or more gaps, depending on the combination of these three parameters. Unfortunately, however, even with constraints on $\hp$ via the assumption of a passively irradiated disk, the turbulent parameter $\alpha$ and the planet's mass $\Mp$ are very difficult to constrain from observations \citep[though some limits on $\alpha$ can be derived, e.g.,][]{dullemond-etal-2018}.

Given that in-plane radiation transport can act as a means of diffusion in that it smears temperature peaks around spiral shocks, its inclusion in a hydrodynamical model could help relax the need to explore very high values of $\alpha$. This helps constrain the parameter space and shift the focus to surveying different planet masses. 

Of course, a cooling prescription raises the question of how one should treat other quantities that become relevant with it, namely the dust opacity and the related properties of dust grains. Progress is constantly being made on this front, with access to up-to-date opacity models for ALMA systems \citep{birnstiel-etal-2018} and tools for opacity-dependent radiative postprocessing of dust and synthetic imaging such as \texttt{RADMC-3D} \citep{dullemond-etal-2012}.

Nevertheless, regardless of the accuracy of the physical model when accounting for cooling, the reality is that the locally isothermal approach will consistently overestimate the planet's ability to open gaps in the inner disk when compared with a model with any form of cooling, as noted by \citet{miranda-rafikov-2019}. This can be translated to the planetary mass estimates of \citet{zhang-etal-2018} simply being a lower limit, and can be leveraged to further constrain the parameter space. 

\subsection{On our approximate recipe for FLD}
\label{sub:approx-fld-disc}

In Sect.~\ref{sub:approx-fld} we defined the parameter $G$ as a proxy for the planet's capacity to open multiple gaps. This was done since we were not interested at the absolute number of rings or gaps that a planet can produce, but rather the relative AMF by planet-driven spiral arms when comparing different equations of state and methods of cooling.

When comparing our models that account for in-plane cooling through FLD (\texttt{rad}) to those that do so with a $\beta$ approach (\texttt{adj}), we found excellent agreement in terms of this metric $G$. However, we also showed that the two methods show somewhat different radial profiles of the angular momentum flux $\FJ$ (Fig.~\ref{fig:amf-adj}), especially far from the planet, as well as very different spiral shock structures due to the diffusion of thermal energy around the shock front in \texttt{rad} models. This difference became visible in our models of ALMA systems in Sect.~\ref{sub:amf-results}, with the gas structure in the inner disk being distinctly different between the two approaches (see Fig.~\ref{fig:systems-gas}). Nevertheless, the differences in gas profiles were ultimately not as noticeable in our dust emission profiles (Fig.~\ref{fig:dust-1D}).

All in all, while this approximate $\beta$ recipe in Eq.~\eqref{eq:badj} is considerably cheaper compared to a model with full radiation transport from a computational point of view ($1.4\times$ for Elias 20, $2.1\times$ for AS 209), we highlight that caution should be exercised when using it.

\subsection{3D effects}
\label{sub:3d-effects}

In this work we approached planet--disk interaction in a 2D, vertically integrated framework. In principle, however, the processes relevant to our study are 3D. Compressed gas can expand in the vertical direction, resulting in an additional degree of freedom around shock fronts and therefore weaker spiral shocks \citep{lyra-etal-2016}. Meridional flows can also slow down or even stall planet-induced gap opening by refilling the gap region \citep{morbidelli-etal-2014,lega-etal-2021}. Finally, the gas can cool and diffuse in the vertical direction, something we incorporate in a simplistic way with $\taueff$ in Eq.~\eqref{eq:Qcool} but that is missed in Eq.~\eqref{eq:Qrad}. The dependence of $\beta$ on height will result in spiral arms cooling at different rates at different altitudes, possibly affecting their AMF. A full 3D investigation with self-consistent global cooling (e.g., an FLD approach) is necessary to compare and contrast our findings with more realistic models. This will be the focus of future work.

\section{Conclusions}
\label{sec:conclusions}

Our aim in this work was twofold: to demonstrate the effects of in-plane radiation transport in planet-driven gap opening, and to compare our findings to those of \mrb, who treated this process using a specially designed local $\beta$ cooling prescription. We approached this topic with nearly-inviscid numerical hydrodynamics simulations of planet--disk interaction.

We first conducted a high-resolution parameter study where we measured the planet-driven spiral angular momentum flux (AMF) as a function of the surface cooling timescale $\bsurf$ and for various treatments of cooling. Namely, we considered the locally isothermal (``\texttt{iso}'') and adiabatic (``\texttt{adb}'') equations of state, as well as models with surface cooling (``\texttt{surf}'') and both surface cooling and in-plane radiation transport via flux-limited diffusion (FLD, ``\texttt{rad}'').

This comparison showed that spiral wave AMF, which can be used as a proxy for the capability of a planet to open one or more gaps, damps fastest when the cooling timescale approaches unity regardless of cooling method. For \texttt{surf} models this translates to $\bsurf\approx 1$, while for \texttt{rad} models the condition is instead $\btot\approx 1.5\text{--}2$ (see Eq.~\eqref{eq:btot}). The former result is consistent with the literature \citep{miranda-rafikov-2020a,zhang-zhu-2020}, while the effect of in-plane radiation transport by means of radiative diffusion is original to this work and shows that a local cooling prescription cannot accurately reproduce models with a full treatment of FLD. We also identified the existence of two regimes, for $\bsurf\lesssim 1$ and $\bsurf\gtrsim1$, where the omission of in-plane cooling results in an under- or overestimation of spiral AMF, respectively.

We then repeated the radiative models described above (\texttt{surf} and \texttt{rad}) using a local $\beta$ relaxation approach instead of allowing the gas to freely heat up and cool down through stellar irradiation and a cooling prescription, respectively. After making necessary adjustments, we showed in Sect.~\ref{sub:approx-fld} that our approximation (Eqs.~\eqref{eq:bmidH}--\eqref{eq:factor-f} for in-plane cooling reproduces our fully radiative models (\texttt{rad}) very accurately, making it a suitable approach when studying a planet's spiral AMF. However, we showed that this $\beta$ approach cannot capture the diffusion of heat around shock fronts, which significantly affects the temperature and density structure of spiral arms and therefore limits the applications of this method.

We followed up with long-term, semi-realistic models of two sources from the DSHARP study \citep{andrews-etal-2018,huang-etal-2018}, namely AS~209 and Elias~20. The two were chosen to probe the two ends of the $\beta$ spectrum (short and long cooling timescales, respectively), in order to verify our findings from the previous numerical experiment (Sect.~\ref{sec:amf-study}). For these models we estimated gas properties using stellar and disk properties from the literature \citep{andrews-etal-2018,zhang-etal-2018} and a temperature-dependent opacity model \citep{lin-papaloizou-1985}.

We found that our models agree very well with our predictions from our parameter study in Sect.~\ref{sec:amf-study}. Namely, the inclusion of in-plane radiation transport results in brighter ring features compared to \texttt{surf} models for AS~209, and weaker gap opening for Elias~20. We then showcased our findings with simplistic synthetic observations of millimeter dust emission for the two systems, showing the presence of an additional (albeit faint) ring at the planet's orbit for AS~209 and much shallower gaps for Elias~20 in \texttt{rad} models.

Our findings highlight the importance of proper treatment of radiation transport in protoplanetary disk modeling, while also showing that the ``single planet, multiple gaps'' scenario demonstrated by \citet{zhang-etal-2018} is certainly possible with realistic gas thermodynamics. Such models can be used to constrain the masses of planets possibly present in the gaps seen in millimeter emission, as well as the turbulence and dust properties in protoplanetary disks (Ziampras et al., in prep.).

\section*{Acknowledgements}
AZ thanks Kees Dullemond for his input and very helpful discussions.
This work was performed using the DiRAC Data Intensive service at Leicester, operated by the University of Leicester IT Services, which forms part of the STFC DiRAC HPC Facility (www.dirac.ac.uk). The equipment was funded by BEIS capital funding via STFC capital grants ST/K000373/1 and ST/R002363/1 and STFC DiRAC Operations grant ST/R001014/1. DiRAC is part of the National e-Infrastructure. This research utilized Queen Mary's Apocrita HPC facility, supported by QMUL Research-IT (http://doi.org/10.5281/zenodo.438045). AZ and RPN are supported by STFC grants ST/P000592/1 and ST/X000931/1, and RPN is supported by the Leverhulme Trust through grant RPG-2018-418. RRR acknowledges financial support through the Ambrose Monell Foundation, and STFC grant ST/T00049X/1. All plots in this paper were made with the Python library \texttt{matplotlib} \citep{hunter-2007}.

\section*{Data Availability}

Data from our numerical models are available upon reasonable request to the corresponding author.

\bibliographystyle{mnras}
\bibliography{refs}

\clearpage

\appendix

\section{Spiral AMF --- resolution study}
\label{apdx:resolution}

\begin{figure}
	\includegraphics[width=\columnwidth]{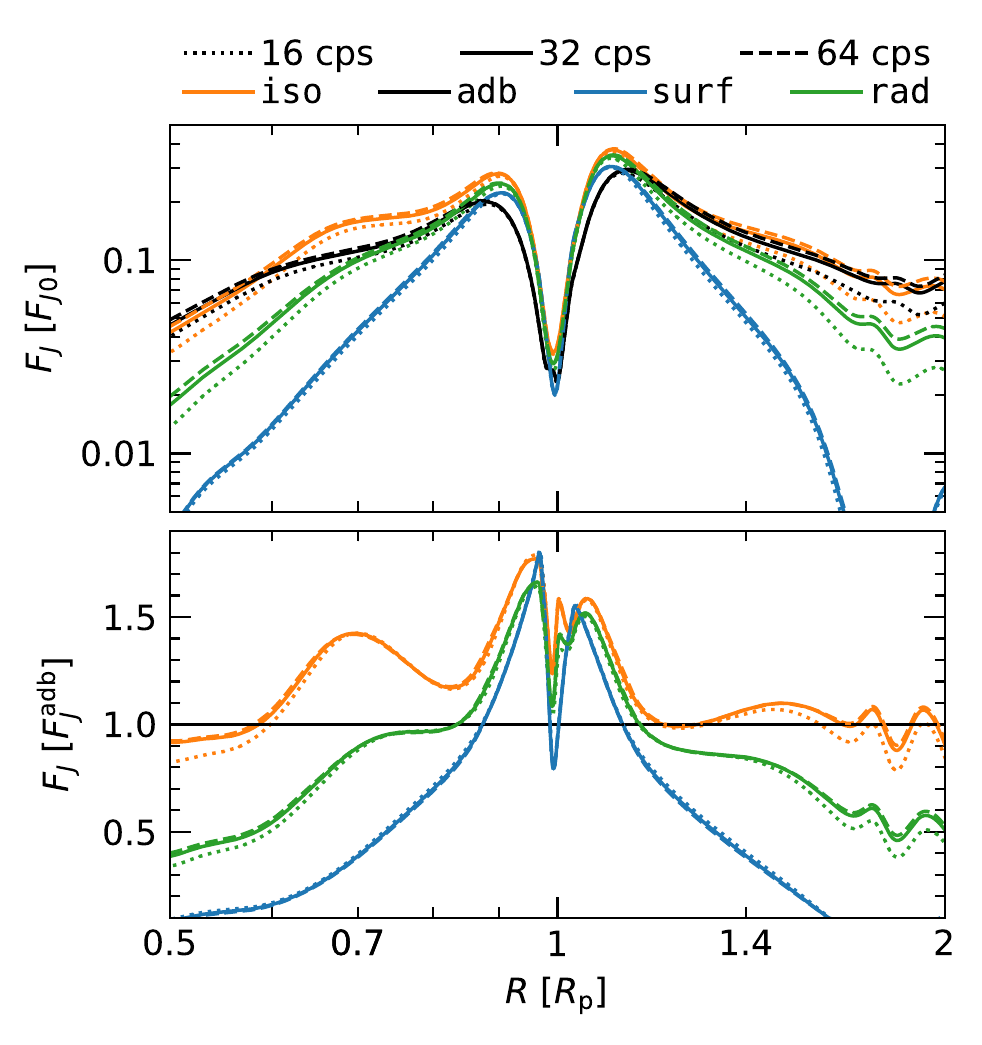}
	\caption{Spiral AMF as a function of resolution in units of cells per scale height at $\Rp$ (cps, various line styles) and equation of state (various colors). Top: $\FJ$ normalized to the reference $\FJj$ in Eq.~\eqref{eq:amf0}. Bottom: $\FJ$ normalized to the curves for adiabatic models (``\texttt{adb}'') at a given resolution.}
	\label{fig:amf-resolution}
\end{figure}

To make sure we resolve the planet-driven spirals accurately enough to calculate their AMF in Sect.~\ref{sec:amf-study}, we carry out a resolution study where we vary the number of cells per scale height (cps) at $\Rp$ from 8 to 64 for the fiducial model with $\tau=10$, $\bsurf=1$ and for various cooling prescriptions. We then compute $\FJ$ for each model, and present our results in Fig.~\ref{fig:amf-resolution}.

We find that a resolution of 32~cps is sufficient for most models, with model \texttt{surf} converging even at 16~cps (top panel of Fig.~\ref{fig:amf-resolution}). There are minor differences between 32 and 64~cps, but the overall behavior is well-resolved. Furthermore, normalizing all curves to $\FJ^\text{adb}$ instead makes any differences between 32 and 64~cps indistinguishable (bottom panel), suggesting that 32~cps is enough to carry out our study in Sect.~\ref{sec:amf-study} while claiming numerical convergence in terms of the parameter $G$. Finally, we note that results at 8~cps were quite poor, and advise against using such low resolutions for similar studies.

\section{Integration of dust particles}
\label{apdx:dust}

To produce the continuum emission maps in Sect.~\ref{sub:dsharp-dust} we closely followed the approach of \mrb. We output azimuthally averaged radial disk profiles (denoted as $\mean{x}$) at a frequency of 10 snapshots per planetary orbit, and then interpolated linearly between these snapshots to generate time-dependent profiles for all quantities. The dust is initialized as $2\times 10^5$ particles of size $a^\text{d}=1$\,mm and bulk density of $\rho^\text{d}=1$\,g/cm${}^3$, distributed radially as $\Sigma_0^\text{d}(R) = 0.01\,\Sigma_0(R)$. The Stokes number is then
\begin{equation}
	\text{St}(R) = \frac{\pi}{2} \frac{\rho_\text{d} a_\text{d}}{\Sigma(R)}.
\end{equation}

We then compute the radial velocity of each dust particle using the dust drift formula by \citet{takeuchi-lin-2002}
\begin{equation}
	\bar{u}_R^\text{d} = \frac{1}{1 + \text{St}^2}\left(\bar{u}_R + \frac{\text{St}}{\mean{\Sigma}\OmegaK}{\D{\mean{P}}{R}}\right),
\end{equation}
where $\bar{u}_R(R) = \mean{\Sigma u_R} / \mean{\Sigma}$ is the density-weighted, azimuthally averaged gas radial velocity profile. The dust particles are then integrated using an RK2 scheme with a timestep of 0.1 orbits, up to a time of 500 planetary orbits. Particles that drift past the inner radial boundary of our domain ($R\leq0.1\Rp$) are dropped.

After that, we compute the dust surface density $\Sigma^\text{d}(R)$ by binning the dust particles in as many bins as there are radial cells in the domain and convolving with a Gaussian curve with a FWHM of $0.075\,R$, similar to \mrb. Finally, we assume a dust opacity of $\kappa^\text{d}=2.2$\,$\text{cm}^2$ per gram of dust and compute the brightness temperature
\begin{equation}
	\Tb(R) = \left(1-e^{-\tau^\text{d}(R)}\right)\,T(R),
\end{equation}
where $\tau^\text{d} = \frac{1}{2}\kappa^\text{d}\Sigma^\text{d}$ and $T(R)$ is the gas temperature. In the optically thin limit, $\Tb$ reflects the surface density distribution of dust grains. 

\section{On the differences between explicit source terms and $\beta$ cooling}
\label{apdx:Q-vs-beta}

In Sect.~\ref{sub:amf-beta-vs-Q} we argued that a $\beta$ cooling approach with $\bsurf$ in the form given by Eq.~\eqref{eq:bsurf} consistently overestimates the true cooling timescale (obtained by explicitly considering $Q_\mathrm{irr}$ and $Q_\mathrm{cool}$) by a factor of 4 for small deviations about the equilibrium temperature. Here, we derive this factor by linearizing the energy equation (Eq.~\eqref{eq:navier-stokes-3}) and using different source terms, similar to \citet{dullemond-etal-2022}.

We start by computing an equilibrium state with temperature $T_0$ given by $\Qirr^0 = \Qcool^0$. We then assume a perturbed temperature $T = T_0 + \delta T$, with a small deviation $\delta T \ll T_0$. When $\beta$ cooling using Eq.~\eqref{eq:bsurf} we have:
\begin{equation}
	\label{eq:cool-using-beta}
	\D{\Sigma e}{t} = Q_\mathrm{relax} = -\frac{T-T_0}{T} \Qcool,\qquad \Qcool = 2\sigmaSB \frac{T^4}{\tauReff},
\end{equation}
see Eqs.~\eqref{eq:Qcool})~\&~\eqref{eq:Qrelax}.
Assuming, for simplicity, that $\tau$ does not depend on temperature, we can now Taylor-expand $\Qrelax$ to first order with respect to $T$ and obtain:
\begin{equation}
	\Qrelax = -2\sigmaSB\frac{(T_0+\delta T)^3}{\tauReff}\delta T \approx -\Qcool^0\frac{\delta T}{T_0}.
\end{equation}

We can repeat this exercise for the case where cooling is determined by
\begin{equation}
	\label{eq:cool-using-deltaQ}
	\D{\Sigma e}{t} = \Qirr - \Qcool = \Delta Q.
\end{equation}
Using Eq.~\eqref{eq:Qirr}, and for a given radial position $R$ and flaring angle $\theta$, we find that $\Qirr$ is independent of temperature. As a result, we can write
\begin{equation}
	\label{eq:deltaQ-Qrelax}
	\begin{split}
		\Delta Q &= \Qirr^0 - \frac{2\sigmaSB}{\tauReff} (T_0 + \delta T)^4 \\
				& = \Qirr^0 - \Qcool^0 - \Qcool^0 \frac{4\delta T}{T_0} = 4 \Qrelax.
	\end{split}
\end{equation}
In other words,  cooling with a $\beta$ approach with $\bsurf$ in the form of Eq.~\eqref{eq:bsurf} will result in small perturbations damping with an effective cooling timescale that is roughly 4 times longer compared to treating cooling with the radiative source terms explicitly. This is verified in Fig.~\ref{fig:G-beta-all}, where curves corresponding to $\beta$ models are offset along the $\beta$ axis, with the offset translating to an effective cooling timescale of a factor 4 times longer for $\beta$ models.

\begin{figure}
	\includegraphics[width=\columnwidth]{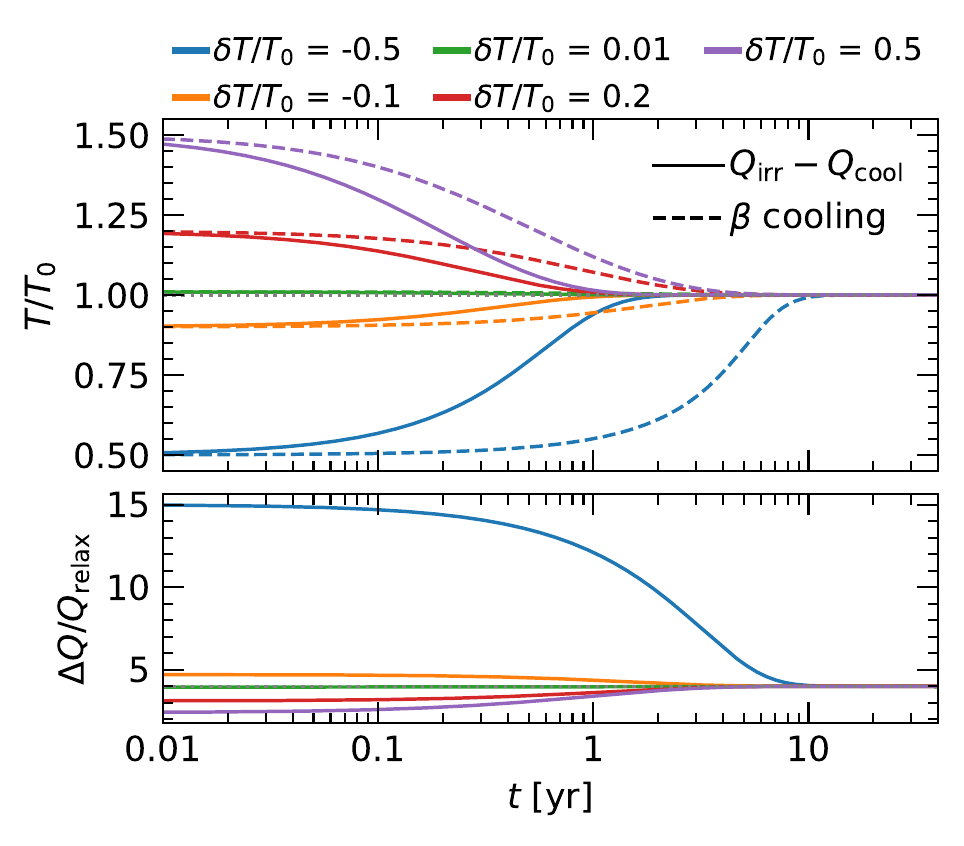}
	\caption{Time evolution of the temperature $T$ for a fixed point in the disk where $T$ evolves via a balance of heating versus cooling (solid curves, $\Delta Q$, Eq.~\eqref{eq:cool-using-deltaQ}) or via $\beta$ relaxation (dashed lines, $\Qrelax$, Eq.~\eqref{eq:cool-using-beta}) to the equilibrium solution $T_0$. The top panel shows that the $\beta$ approach consistently overestimates the true cooling timescale. The bottom panel shows that the ratio $\Delta Q/\Qrelax$ converges to a value of 4 for small deviations from $T_0$ (see Eq.~\eqref{eq:deltaQ-Qrelax}) but can be substantially different for larger deviations.}
	\label{fig:Q-vs-beta-timeevo}
\end{figure}

We also note that this factor of 4 can vary significantly as perturbations grow in strength. To showcase this, we integrate Eqs.~\eqref{eq:cool-using-beta}~\&~\eqref{eq:cool-using-deltaQ} in time with a simple \texttt{python} script for various values of $\delta T/T_0$ and consolidate our findings in Fig.~\ref{fig:Q-vs-beta-timeevo}. As shown in this figure, this approach is no longer valid for $|\delta T|/T_0\gtrsim0.1$, making the $\beta$ approach unsuitable for modeling massive planets in disks subject to shock heating \citep{rafikov-2016,ziampras-etal-2020a}. In particular, a $\beta$ cooling approach results in a completely different relaxation rate for $\delta T/T_0 \lesssim -0.5$.

Finally, it should be noted that for a temperature-dependent opacity this approximation becomes significantly more complicated. \citet{dullemond-etal-2022} showed that this factor of 4 is more generally $4+b$ in the optically thin limit (and therefore $4-b$ in the optically thick limit), where $b = \D{\log\kappa}{\log T}$. In the interest of only highlighting the shortcomings of $\beta$ cooling, we chose not to repeat models that explore different values of $b$.

\FloatBarrier

\section{Nonlinear regime}
\label{apdx:nonlinear}

In the majority of our runs we use a planet mass $\Mp=0.3\,\Mth$ such that our models probe the (quasi-)linear regime of planet--disk interaction. For more massive planets this interaction becomes nonlinear, with the planet launching multiple strong spiral shocks that result in a more complex disk structure due to the opening of multiple deep gaps and stronger shock heating \citep{rafikov-2016,ziampras-etal-2020a}. The latter in particular can lead to a temperature increase of up to a factor of 2 from the equilibrium solution for $\Mp\gtrsim\Mth$ in optically thick disks \citep{ziampras-etal-2020a}.

This situation creates two problems in the scope of our study. For one, the excitation of multiple spiral shocks causes the AMF $\FJ(R)$ to show multiple peaks in the inner disk, making the parameter $G$ less effective at characterizing a planet's capacity to open multiple gaps. At the same time, large deviations from the equilibrium temperature $T_0$ will change the correction factor of 4 derived for the linear regime in Appendix~\ref{apdx:Q-vs-beta} in a non-trivial way, which could render $\beta$ models less suitable for the treatment of cooling. To investigate the above, we carry out a set of models with $\tau=10$, $\hp=0.05$ and a higher mass planet with $\Mp=3\,\Mth$ and compute once again $G(\bsurf)$.

\begin{figure}
	\includegraphics[width=\columnwidth]{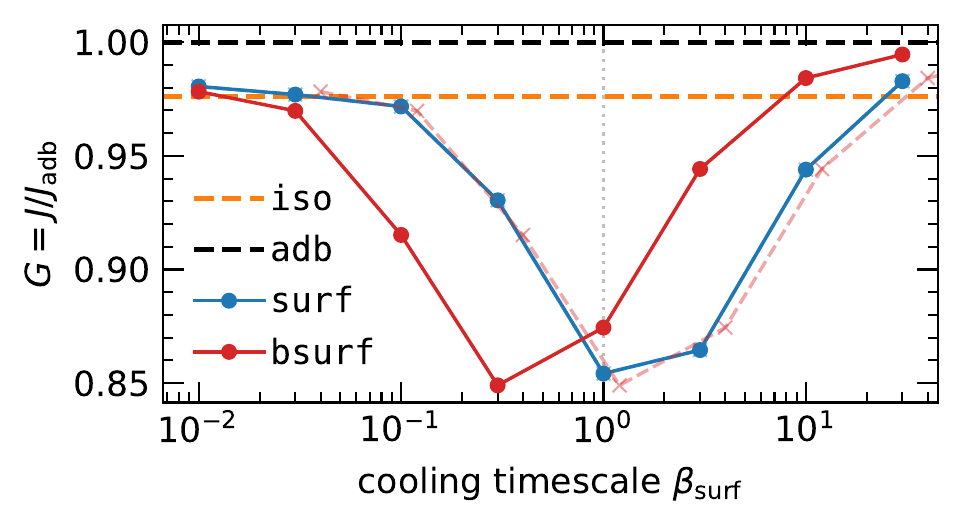}
	\caption{The parameter $G$ similar to Fig.~\ref{fig:G-beta}, but for models with $\Mp=3\,\Mth$. We find a very good match between models \texttt{bsurf} and \texttt{surf} after shifting by the correction factor of 4 discussed in Appendix~\ref{apdx:Q-vs-beta}, even in this nonlinear regime.}

	\label{fig:G-beta-Mp3}
\end{figure}

Our results are shown in Fig.~\ref{fig:G-beta-Mp3}. We find that shifting \texttt{bsurf} models to the right by a correction factor of 4 yields a very good match with the \texttt{surf} models, albeit the optimal factor would be $\approx 3.5$. This is in line with our estimate that this correction factor should decrease for larger $\delta T/T_0$ (stronger shock heating) in Fig.~\ref{fig:Q-vs-beta-timeevo}. Given that a planet mass of $3\,\Mth$ is already quite high, we conclude that a correction factor of 4 will be sufficient in most cases.

At the same time, however, the data becomes quite noisy for this planet mass. The planet strongly influences its surrounding disk, with the surface density changing locally by a factor of up to 2--3 depending on equation of state. This is in stark contrast to our models with $\Mp=0.3\,\Mth$, where $\Sigma$ barely changes by 20\% in most cases (see Fig.~\ref{fig:fiducial-planet-sig}), and complicates the interpretation of $G$ as a measure of the planet's capacity to open multiple gaps.

We also find that, here, $G^\text{adb}\approx G^\text{iso}$. This is not unexpected, as for such high $\Mp$ the density waves are highly nonlinear and dissipate through shocks immediately around the planet, before the linear propagation of the spiral can change their AMF.

\FloatBarrier

\section{Effect of the aspect ratio on spiral AMF}
\label{apdx:aspect-ratio}

\begin{figure}
	\includegraphics[width=\columnwidth]{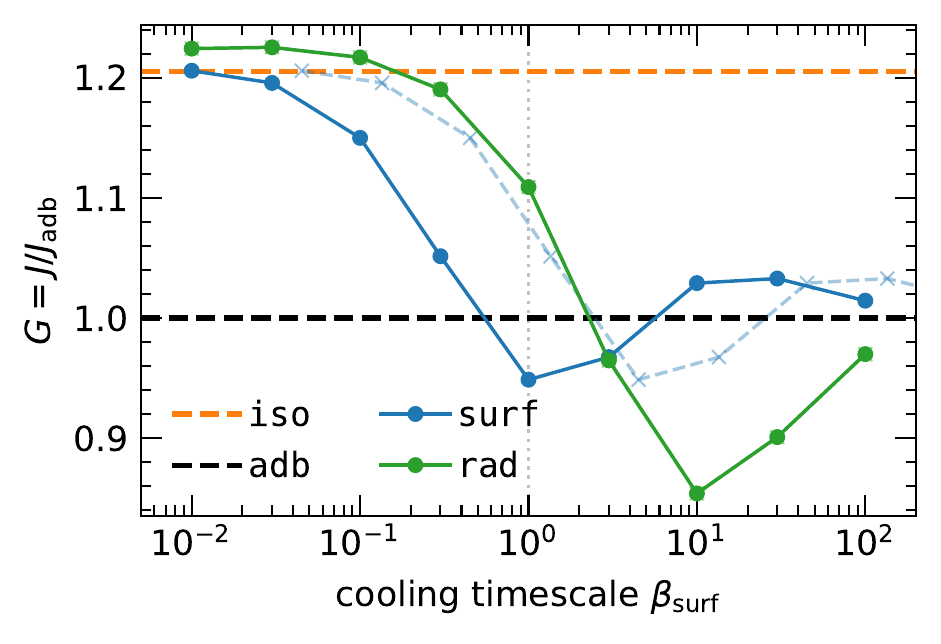}
	\caption{The parameter $G$ similar to Fig.~\ref{fig:G-beta}, but for models with $\hp=0.1$. The dashed blue line results from shifting the \texttt{surf} model by $1+f$ (see Eq.~\eqref{eq:factor-f}).}
	\label{fig:G-beta-h1}
\end{figure}

In all models shown in the main text we used a reference aspect ratio $\hp=0.05$. It is nevertheless worth investigating whether the critical cooling timescale $\beta_\text{crit}$ for which $G$ is smallest might depend on $h$. To answer this question, we carry out a set of models with $\tau=10$ and $\hp=0.1$ (with $\Mp=0.3\,\Mth=10^{-4}\,\Mstar$) and plot the computed $G(\bsurf)$ in Fig.~\ref{fig:G-beta-h1}. As shown there, while the absolute value of $G$ is typically higher compared to models with $\hp=0.05$, $G$ once again shows a minimum at $\bsurf=1$ and 10 for models \texttt{surf} and \texttt{rad}, respectively, consistent with our findings for $\hp=0.05$. The values of $\Sigma$ and $\kappa$ used in these models are listed in Table~\ref{table:amf-values-2}.

\FloatBarrier

\section{Comparison with recipe of MR20b}
\label{apdx:mr20b}

\begin{figure}
	\includegraphics[width=\columnwidth]{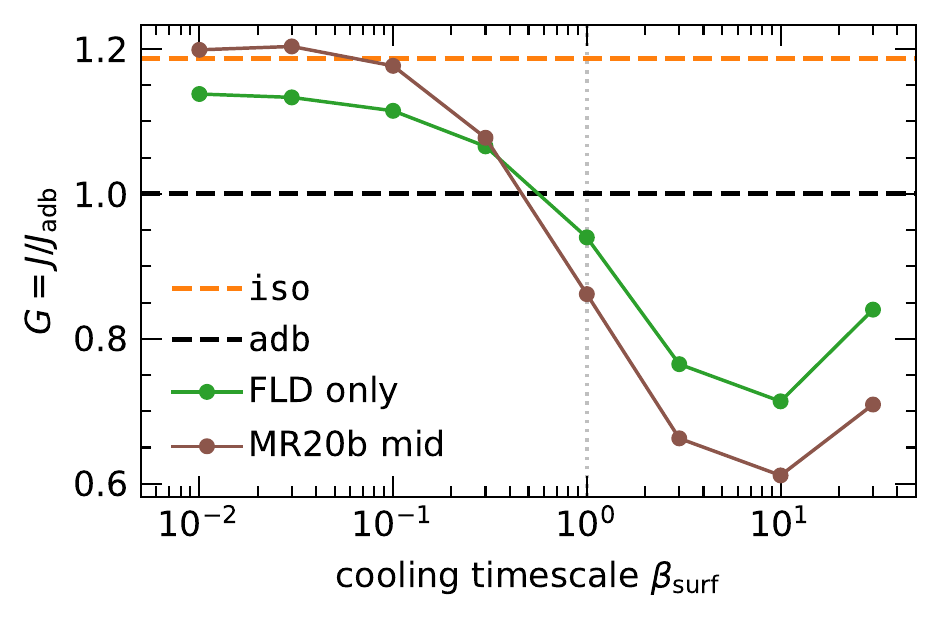}
	\caption{The parameter $G(\bsurf)$ for $\tau=10$ similar to Fig.~\ref{fig:G-beta}, comparing models where only in-plane cooling is considered. Brown: $\beta$ models with the in-plane cooling recipe given by Eqs.~\eqref{eq:bmid}~\&~\eqref{eq:kstar}, following \mrb. Green: models with full in-plane radiative diffusion ($\Qrad$, Eq.~\eqref{eq:Qrad}).}
	\label{fig:G-beta-mr20b}
\end{figure}

\begin{figure}
	\includegraphics[width=\columnwidth]{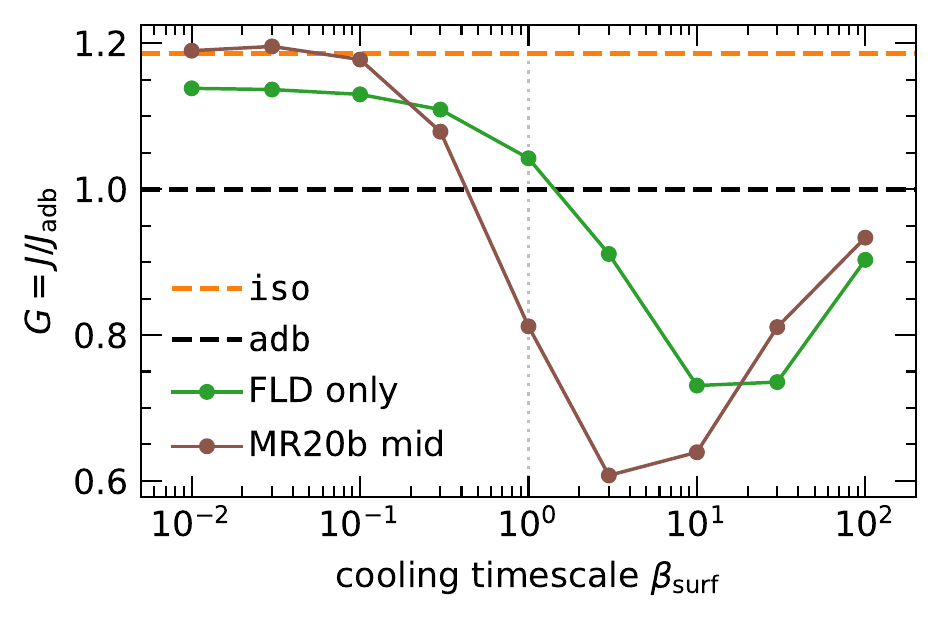}
	\caption{Similar to Fig.~\ref{fig:G-beta-mr20b} for $\tau=1$.}
	\label{fig:G-beta-mr20b-tau1}
\end{figure}

\begin{figure}
	\includegraphics[width=\columnwidth]{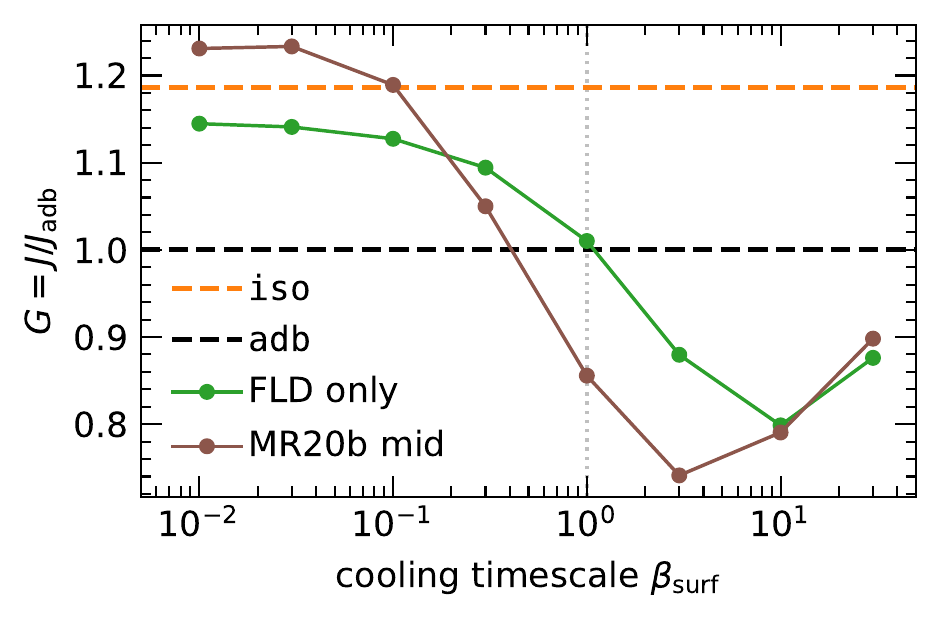}
	\caption{Similar to Fig.~\ref{fig:G-beta-mr20b} for $\hp=0.1$.}
	\label{fig:G-beta-mr20b-h1}
\end{figure}

We also discuss the performance of the effective $\beta$-cooling prescription proposed in \citetalias{miranda-rafikov-2020b} (see Eqs.~\eqref{eq:bmid}~\&~\eqref{eq:kstar}). In Figs.~\ref{fig:G-beta-mr20b}--\ref{fig:G-beta-mr20b-h1} we compare the performance of just the $\bmid$ part of their prescription (marked ``MR20b mid'') against models with only $\Qrad$ (i.e., no surface cooling or irradiation heating are included), marked "FLD only". This enables a direct comparison of the two ways to treat the in-plane cooling. Different plots make different assumptions about $\tau$ and $\hp$, as indicated in captions. One can see that the simple prescription of \citetalias{miranda-rafikov-2020b} does not do very well at reproducing the direct radiation diffusion calculations: it is often offset horizontally and features a deeper minimum than the latter.

\begin{figure}
	\includegraphics[width=\columnwidth]{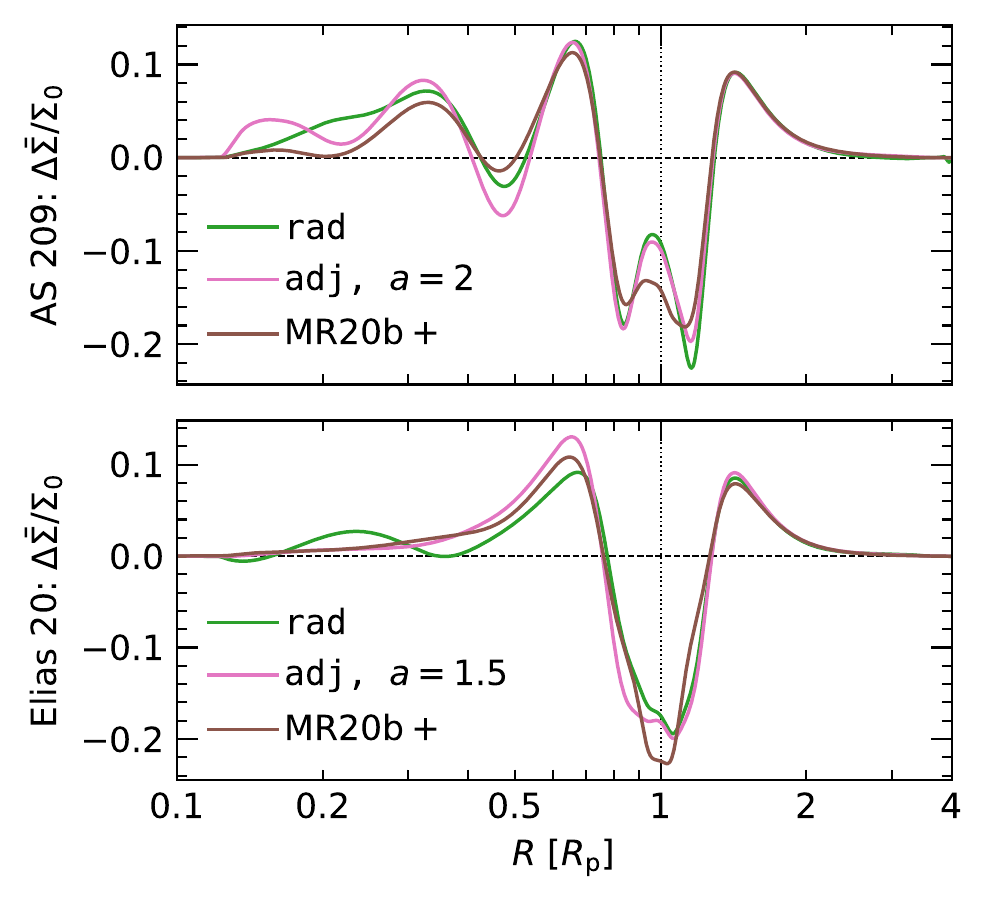}
	\caption{Azimuthally averaged gas density after 500 planetary orbits for AS~209 and Elias~20, similar to Fig.~\ref{fig:systems-gas}. }
	\label{fig:systems-gas-mr20b}
\end{figure}

In Fig.~\ref{fig:systems-gas-mr20b} we also show the surface density profiles obtained for AS 209 and Elias 20 using different approximations of surface and in-plane cooling. Here, models marked ``MR20b+'' use $\beta$ cooling as in Eq.~\eqref{eq:btot} where $\bsurf$ is corrected by the factor 4 discussed in Sec.~\ref{sub:amf-beta-vs-Q} and $\bmid$ is given by the recipe in \citetalias{miranda-rafikov-2020b} through Eqs.~\eqref{eq:bmid}~\&~\eqref{eq:kstar}. One can see that the latter approach gives results different from both the adjusted (\texttt{adj}) and most realistic models (\texttt{rad}) when it comes to details around rings/gaps.

The likely reason for the poor match between the \citetalias{miranda-rafikov-2020b} and the \texttt{rad} models is that the former effectively assumed the characteristic radial scale of the temperature perturbation in the planet-driven density wave to be the same as the radial scale of its $\Sigma$ perturbation, which becomes considerably smaller than $H$ as the wave propagates. However, as Fig.~\ref{fig:rad-corr-shock} clearly illustrates, thermal diffusion ahead of the shock makes the thermal length scales larger than the length scale of $\Sigma$ variation, which acts to slow down the loss of the wave AMF compared to  \citetalias{miranda-rafikov-2020b}. In fact, one can argue that in real disks the temperature length scale in the disk plane should always be around $H$: once the photons diffuse by $\sim H$ horizontally, it becomes easier (more likely) for them to then diffuse vertically by $\sim H$ than to continue horizontally. This may justify the better performance of the simple prescription (Eq.~\eqref{eq:bmidH}) in reproducing \texttt{rad} models upon the simple adjustment described in Sect.~\ref{sub:approx-fld}.

\FloatBarrier

\section{Values of $\Sigma$ and $\kappa$ in our AMF study}
\label{apdx:tables}

Here we provide a summary of the values of $\Sigma$ and $\kappa$ used to achieve different combinations of $\bsurf$ and $\tau$ in Sect.~\ref{sec:amf-study}. We note that some values or combinations of $\Sigma$ and $\kappa$ are unphysical, but this does not matter in the scope of our study as the intent is simply to achieve a given combination of $\bsurf$ and $\tau$. Our parameters are listed in Table~\ref{table:amf-values}.

\begin{table}
	\begin{center}	
		\parbox{\textwidth}{\caption{Values for surface density $\Sigma$ and opacity $\kappa$ used in our models in Sect.~\ref{sec:amf-study}, and the corresponding surface cooling timescale $\bsurf$ and optical depth $\tau$. Alongside an aspect ratio profile $h = 0.05\,(R/\Rp)^{1/4}$, or equivalently a temperature profile $T\propto R^{-1/2}$, our choices ensure constant $\bsurf$ and $\tau$ throughout the disk.}}
		\label{table:amf-values}
		\begin{tabular}{c | c | c | c | c | c | c | c | c | c}
			\hline
			quantity & \multicolumn{9}{c}{value} \\\hline
			$\tau$ & \multicolumn{9}{c}{10} \\\hline
			$\bsurf$ & 0.01 & 0.03 & 0.1 & 0.3 & 1 & 3 & 10 & 30 &100 \\\hline
			$\Sigma$ [$\mathrm{g}\,\mathrm{cm}^{-2}$] & 3.8146 & 11.552 & 37.49 & 113.54 & 377.09 & 1141.9 & 3792.7 & 11367 & 37892 \\\hline
			$\kappa$ [$\mathrm{cm}^2\,\mathrm{g}^{-1}$] & 5.2699 & 1.7103 & 0.5362 & 0.17503 & 0.05362 & 0.017402 & 0.0053311 & 0.0017594 & $5.278\times10^{-4}$ \\\hline\hline
			quantity & \multicolumn{9}{c}{value} \\\hline
			$\tau$ & \multicolumn{9}{c}{1} \\\hline
			$\bsurf$ & 0.01 & 0.03 & 0.1 & 0.3 & 1 & 3 & 10 & 30 & 100 \\\hline
			$\Sigma$ [$\mathrm{g}\,\mathrm{cm}^{-2}$] & 15.151 & 45.883 & 148.91 & 450.96 & 1506.4 & 4588.3 & 14891.0 & 45211 & 150710 \\\hline
			$\kappa$ [$\mathrm{cm}^2\,\mathrm{g}^{-1}$] & 0.13192 & 0.044067 & 0.0135 & 0.0043813 & 0.0013345 & $4.356\times10^{-4}$ & $1.335\times10^{-4}$ & $4.4236 \times 10^{-5}$ & $1.33 \times 10^{-5}$ \\\hline
		\end{tabular}
	\end{center}
	
\end{table}

\begin{table}
	\begin{center}
		\parbox{\textwidth}{\caption{Values for surface density $\Sigma$ and opacity $\kappa$ used in our models in Appendix~\ref{apdx:aspect-ratio}, where $\hp=0.1$, similar to Table~\ref{table:amf-values}.}}
		\label{table:amf-values-2}
		\begin{tabular}{c | c | c | c | c | c | c | c | c | c}
			quantity & \multicolumn{9}{c}{value} \\\hline
			$\tau$ & \multicolumn{9}{c}{10} \\\hline
			$\bsurf$ & 0.01 & 0.03 & 0.1 & 0.3 & 1 & 3 & 10 & 30 &100 \\\hline
			$\Sigma$ [$\mathrm{g}\,\mathrm{cm}^{-2}$] & 242.5 & 727.51 & 2425.1 & 7275.1 & 24250 & 72751 & $2.425\times 10^5$ & $7.275\times10^5$ & $2.43\times10^6$\\\hline
			$\kappa$ [$\mathrm{cm}^2\,\mathrm{g}^{-1}$] & 0.082473 & 0.027491 & 0.0082472 & 0.0027491 & $8.247\times10^{-4}$ & $2.749\times10^{-4}$ & $8.247\times 10^{-5}$ & $2.749\times 10^{-5}$ & $8.247\times 10^{-6}$\\\hline
		\end{tabular}
	\end{center}
\end{table}

\bsp	
\label{lastpage}
\end{document}